\documentclass[journal]{IEEEtran}
\usepackage{amsfonts}
\usepackage{times}
\usepackage{graphicx}
\usepackage{latexsym}
\usepackage{dsfont}
\usepackage{amssymb}
\usepackage{amsmath}
\usepackage{cite}
\usepackage{verbatim}

\newtheorem{algorithm}{Algorithm}

\newcommand{\figref}[1]{{Fig.}~\ref{#1}}

\def\bb0{{\mathbb{0}}}

\def\ba{{\mathbf{a}}}
\def\bb{{\mathbf{b}}}
\def\bc{{\mathbf{c}}}

\def\bff{{\mathbf{f}}}
\def\bg{{\mathbf{g}}}
\def\bh{{\mathbf{h}}}

\def\bm{{\mathbf{m}}}
\def\bn{{\mathbf{n}}}

\def\bu{{\mathbf{u}}}

\def\bw{{\mathbf{w}}}
\def\bx{{\mathbf{x}}}
\def\by{{\mathbf{y}}}
\def\bz{{\mathbf{z}}}
\def\b0{{\mathbf{0}}}

\def\bA{{\mathbf{A}}}

\def\bF{{\mathbf{F}}}
\def\bG{{\mathbf{G}}}
\def\bH{{\mathbf{H}}}
\def\bI{{\mathbf{I}}}

\def\bR{{\mathbf{R}}}

\def\bbC{{\mathbb{C}}}

\def\bbE{{\mathbb{E}}}

\def\bbR{{\mathbb{R}}}

\def\cA{\mathcal{A}}

\def\cC{\mathcal{C}}

\def\cF{\mathcal{F}}
\def\cG{\mathcal{G}}

\def\cN{\mathcal{N}}

\def\cV{\mathcal{V}}

\def\sf0{{\mathsf{0}}}

\usepackage{epstopdf}
\usepackage{enumerate}
\usepackage{amsmath}
\usepackage{mathtools}
\usepackage{float}
\usepackage{color}
\usepackage{makeidx}
\usepackage{bm}
\usepackage{cleveref}
\usepackage{url}
\usepackage{steinmetz}
\usepackage{varwidth}
\usepackage{soul}
\usepackage{makecell}
\usepackage{booktabs}
\usepackage{comment}
\usepackage{bbm}
\usepackage{subfigure}
\usepackage{balance}

\let\oldFootnote\footnote
\newcommand\nextToken\relax
\renewcommand\footnote[1]{%
	\oldFootnote{#1}\futurelet\nextToken\isFootnote}
\newcommand\isFootnote{%
	\ifx\footnote\nextToken\textsuperscript{,}\fi}

\newcommand{\sref}[1]{{Section}~\ref{#1}}

\usepackage[normalem]{ulem}
\newcommand{\tdel}[1]{{\color{red}\bgroup\markoverwith{\rule[0.5ex]{3pt}{0.5pt}\hskip2pt}\ULon{#1}}}

\DeclareMathOperator\RE{\mathrm{Re}}

\DeclarePairedDelimiter{\ceil}{\lceil}{\rceil}

\newcommand*{\E}{\mathrm{e}}

\def \rm {\mathrm}

\usepackage{algorithm}
\makeatletter
\newcommand{\algorithmname}{\ALG@name}
\renewcommand{\floatc@ruled}[2]{{\@fs@cfont #1:} #2\par}
\makeatother
\usepackage[noEnd=false]{algpseudocodex}
\tikzset{algpxIndentLine/.style={draw=black}}
\algrenewcommand{\alglinenumber}[1]{\bfseries\footnotesize #1}
\algrenewcommand{\textproc}{}
\algrenewcommand{\algorithmicrequire}{\textbf{Initialization:}}  
\algrenewcommand{\algorithmicensure}{\textbf{Output:}} 

\begin{document}
\bstctlcite{IEEEexample:BSTcontrol}

\title{ISAC with Backscattering RFID Tags: \\ Beamforming and Codebook Design}

\author{
	\IEEEauthorblockN{Hao Luo, Umut Demirhan, and Ahmed Alkhateeb}
    \thanks{Part of this work was presented at the IEEE International Conference on Communications (ICC) 2024 \cite{Luo2024}.}
    \thanks{Hao Luo, Umut Demirhan, and Ahmed Alkhateeb are with the Wireless Intelligence Lab and the School of Electrical, Computer, and Energy Engineering at Arizona State University (email: \{h.luo, udemirhan, alkhateeb\}@asu.edu). This work was supported in part by the National Science Foundation (NSF) under Grant No. 2229530.}
}

\maketitle

\begin{abstract}	
	This paper explores an integrated sensing and communication (ISAC) system with backscattering radio frequency identification (RFID) tags. In this setup, an access point employs communication beams to serve communication users while leveraging a sensing beam to interrogate RFID tags. Under the total transmit power constraint of the system, our objective is to design a joint sensing and communication beamforming codebook by considering the tag interrogation and communication requirements. To lay a foundation for the codebook design problem, we first study the beamforming design problem in the single-tag scenario and investigate two approaches: a two-stage zero-forcing-based design with optimized power allocation and a joint beamforming design that minimizes the total transmit power. Extending our study to the multi-tag scenario, we propose a site-specific beamforming codebook design. The proposed approach partitions the region of interest into angular sectors and jointly optimizes the sensing and communication beams for each sector. The simulation results show that the proposed beamforming designs can effectively mitigate the mutual interference between sensing and communication functionalities, thus enhancing the interrogation range of the tags while minimizing transmit power. Also, the efficacy of the proposed site-specific codebook design has been demonstrated in terms of interrogation success rate, offering a promising approach for ISAC backscattering systems.
\end{abstract}


\section{Introduction} \label{sec:introduction}
Integrated sensing and communication (ISAC) \cite{Liu2022a,Demirhan2023a} is emerging as a key enabler for next-generation wireless networks, aiming to provide enhanced spectral, energy, and hardware efficiency. A potential application of this paradigm is the use of radio frequency identification (RFID) for inventory management in warehouses or retail stores, where low-cost passive RFID tags replace conventional barcodes \cite{Athauda2018,Saetia2025}.
In such scenarios, an ISAC system can interrogate RFID tags attached to goods for identification and tracking while simultaneously transmitting downlink data to communication users, such as wireless surveillance cameras or mobile devices, in a cooperative manner.
However, this integration presents several challenges: (i) passive tags have a limited reading range due to the absence of a built-in power source; (ii) sensing and communication signals may mutually interfere, degrading tag-reading reliability; (iii) the exact tag positions and the associated channel information are usually unknown. To address the first two challenges, joint multiple-input multiple-output (MIMO) beamforming design for sensing and communication is essential to enhance the interrogation range and suppress mutual interference. Furthermore, the third challenge calls for a codebook of beams and a scanning strategy to discover and interrogate tags with unknown positions and channel information. Motivated by these needs, this paper studies beamforming and codebook design for ISAC systems with backscattering RFID tags.

\subsection{Related Work}
\subsubsection{ISAC Beamforming Design}
MIMO beamforming is an effective technique to mitigate the mutual interference between sensing and communication. By leveraging spatial degrees of freedom, transmit beamforming \cite{Liu2020,Liu2022b,Barneto2022,Hua2023} steers independent beams toward communication users and sensing targets, thereby reducing the mutual interference. For instance, in \cite{Liu2020}, the authors proposed a precoder design that synthesizes desired sensing patterns while satisfying communication signal-to-interference-plus-noise ratio (SINR) and power constraints. Furthermore, recent studies \cite{Zhao2022,Ni2022} have explored joint transmit and receive beamforming via alternating optimization algorithms. In the realm of codebook design, researchers have developed specialized codebooks \cite{Bayraktar2023,Hernangomez2026} to minimize self-interference in full-duplex ISAC scenarios. However, existing works address general ISAC beamforming without considering the integration of RFID tags, which have distinct requirements compared to conventional sensing targets.

\subsubsection{MIMO Beamforming for RFID Systems}
The reading range of passive RFID tags is typically limited due to the lack of an internal power source, and MIMO beamforming can extend it by exploiting the beamforming gain of multiple antennas. In \cite{Chen2016}, the authors introduced a blind adaptive beamforming algorithm for multiple-antenna RFID systems, improving the interrogation range of the tags and the data transmission performance. In \cite{Wang2019}, the authors leveraged a distributed MIMO system and designed a blind distributed beamforming algorithm to further enhance the interrogation range. To increase the tag reading rate, the authors in \cite{Wang2023,Pirayesh2023} explored concurrent tag reading, where the RFID reader employs beamforming to interrogate multiple tags simultaneously. However, these studies primarily focused on RFID systems without considering the coexistence of a communication function, which is a key aspect of ISAC systems.

\subsubsection{Joint Beamforming for ISAC with Backscattering Tags}
The integration of ISAC with backscattering tags was first proposed in \cite{Galappaththige2023}, where a base station employs transmit beamforming to communicate with a user while broadcasting sensing signals to interrogate an RFID tag. This study analyzed the communication performance of both the user and the backscattering tag, as well as the sensing performance at the base station, under different power allocation strategies between sensing and communication. However, it did not consider the interrogation requirements of RFID tags, specifically the sensitivities of both the tag and the reader, which are crucial in practical applications. Additionally, it did not apply transmit beamforming to sensing signals, which is vital for extending the RFID interrogation range. 
To address these limitations, our preliminary work \cite{Luo2024} investigated the zero-forcing-based and joint beamforming designs while accounting for the sensitivity requirements of both the tag and the reader. This work considered a single RFID tag and a single communication user, and it assumed that the tag channel information was known to the access point.

Following these preliminary efforts, various other studies have explored the joint beamforming design for ISAC systems with backscattering tags~\cite{Zhao2024, Rojith2025a, Rojith2025b,Zargari2025}. For example, in \cite{Zhao2024}, the authors proposed a backscatter-ISAC system and developed a successive convex approximation based joint beamforming scheme to maximize the communication rate under strict sensing performance constraints and power budgets. In \cite{Zargari2025}, the authors introduced an integrated sensing and backscatter communication framework and employed an alternating optimization algorithm to jointly design the beamformers, sensing signals, sensing combiners, and reflection coefficients with the objective of minimizing the total base station transmit power. 
These studies \cite{Luo2024,Zhao2024,Rojith2025a,Rojith2025b,Zargari2025}, however, did not consider the unknown positions or channel information of the tags and the associated codebook design problem and scanning strategy, both of which are critical aspects in practical RFID applications. To address this gap, we take advantage of site-specific channel data collected offline using reference tags placed at certain locations, which serve only as spatial sampling points for codebook construction. The resulting codebook can therefore be designed prior to deployment and used to interrogate tags at arbitrary locations within the region of interest without prior knowledge of their locations or channels.

\subsection{Contributions}
 To the best of our knowledge, this is the first work to study joint sensing and communication beamforming and codebook design for RFID-enabled ISAC while accounting for both the RFID interrogation and communication requirements. The main contributions are summarized as follows:
\begin{itemize}
	\item \textbf{Formulating an ISAC system model with backscattering RFID tags and a beamforming codebook design problem.} We consider a multi-antenna access point that interrogates passive RFID tags while simultaneously serving multiple communication users. We develop a signal model that accounts for the RFID command and tag-reply intervals and the mutual interference among the interrogation, communication, and backscattered signals. In particular, we incorporate the tag-side and reader-side SINR requirements for successful interrogation, along with the SINR requirements of the communication users. Under a total transmit power constraint, we then formulate the beamforming codebook design problem to maximize the number of successfully interrogated tags over a complete beam sweep.
	\item \textbf{Developing two beamforming design solutions for the single-tag scenario.} We first develop a two-stage zero-forcing approach that constructs the sensing and communication beam directions to eliminate direct inter-beam interference and subsequently optimizes their power allocation. The resulting power allocation problem is a linear program, based on which we identify the dominant RFID SINR constraint and derive the closed-form solution. Second, we propose a joint beamforming design that jointly optimizes the sensing and communication beamformers by minimizing the total transmit power subject to the tag-side, reader-side, and communication user SINR requirements. We reformulate the SINR constraints as equivalent second-order-cone constraints, allowing the joint design to be efficiently solved as a convex program.
	\item \textbf{Developing a site-specific beamforming codebook design for the multi-tag scenario.} We relate the expected number of successfully interrogated tags to spatial interrogation coverage and propose a sector-based design with one joint sensing and communication codeword per angular sector. We sample each sector using a grid of reference locations and construct an offline database from their channel measurements. Each codeword is optimized to maximize the number of covered reference locations subject to the communication user SINR and transmit power requirements, yielding a mixed-integer problem solved using semidefinite relaxation and generalized Benders decomposition. The grid serves only for spatial sampling, and the designed codebook applies to tags at arbitrary locations within the region of interest.
\end{itemize}
Extensive simulations demonstrate that both proposed beamforming solutions effectively mitigate mutual interference and extend the interrogation distance. Moreover, the proposed codebook substantially outperforms the benchmark methods, approaches the upper bound with a fine sector resolution, and remains effective with multiple communication users and under multipath propagation.

The remainder of this paper is organized as follows. Sections~\ref{sec:system_model} and~\ref{sec:problem_formulation} present the system model and problem formulation. \sref{sec:beamforming_design_single_tag} develops the single-tag beamforming solutions, and \sref{sec:codebook_design_multi_tag} presents the multi-tag codebook design. \sref{sec:simulation_results} provides simulation results, and \sref{sec:conclusion} concludes the paper.

\textbf{Notation:}
$\bA$ is a matrix; $\ba$ is a vector; and $a$ is a scalar. $\cA$ is a set. 
$\bA^T$ and $\bA^H$ are the transpose and Hermitian transpose of $\bA$, respectively. 
$[\ba]_i$ denotes the $i^\rm{th}$ element of the vector $\ba$.
$\|\ba\|$ denotes the 2-norm of the vector $\ba$.
$\rm{Tr}(\bA)$ denotes the trace of the matrix $\bA$.
$\cC\cN(\cdot,\cdot)$ denotes a complex Gaussian distribution, whose first and second arguments specify the mean and covariance, respectively. The mean may be scalar- or vector-valued, while the covariance may be scalar- or matrix-valued, as appropriate.
$\bbE[\cdot]$ is used to denote expectation.

\section{System Model} \label{sec:system_model}

\begin{figure}[!t]
	\centering	
    \includegraphics[width=0.825\columnwidth]{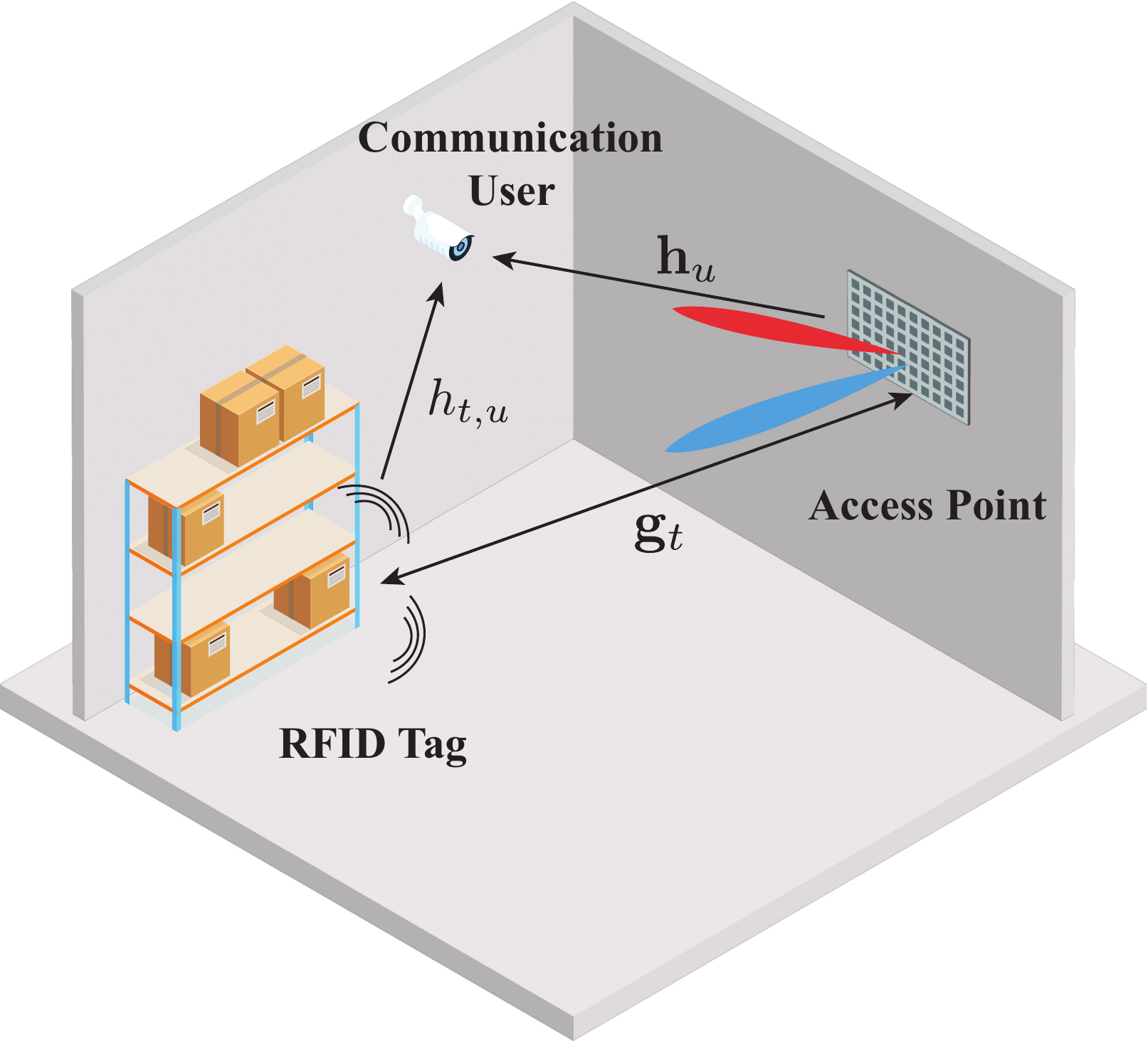}
	\caption{Illustration of the MIMO ISAC system where an access point transmits signals for both user communication and passive RFID tag interrogation. This figure depicts a potential application for inventory management and surveillance in a warehouse or retail store.}
	\label{fig:system_model}
\end{figure}

We consider a MIMO ISAC system comprising an access point, $T$ passive RFID tags, indexed by $t\in\mathcal{T}=\{1,2,\ldots,T\}$, and $U$ static communication users, indexed by $u\in\mathcal{U}=\{1,2,\ldots,U\}$, as illustrated in \figref{fig:system_model}.
Hereafter, communication refers to the direct downlink transmission of user data, whereas sensing refers to RFID interrogation, including backscattered tag replies. Accordingly, the access point simultaneously transmits an RFID-interrogation signal and communication signals intended for the users.\footnote{The access point is equipped with RFID reader functionality. The terms ``access point'' and ``reader'' are used interchangeably throughout this paper.}
Following standard RFID protocols~\cite{EPCGlobal}, we assume tag replies are singulated and decoded one at a time through an anti-collision procedure. During its assigned reply slot, a tag modulates the incident sensing signal to backscatter its stored information, which the access point can process for sensing tasks such as identification and tracking~\cite{Xu2023}. The access point employs a monostatic configuration where the transmitter and receiver share a common $M$-element antenna array via self-isolation circuitry (e.g., circulators or directional couplers). We assume sufficient isolation is achieved to suppress self-interference effectively~\cite{Boaventura2016} and focus on modeling the signals received at the tag, the communication users, and the reader.

\subsection{Preliminaries on Passive RFID Interrogation} \label{subsec:rfid_interrogation}
 We consider the EPCglobal Class-1 Generation-2 (Gen2) UHF air interface~\cite{EPCGlobal}, which is widely used in commercial RFID systems. A passive tag has no internal power source and obtains its operating energy from the signal radiated by the reader. Tag interrogation proceeds through a sequence of protocol operations. During a command interval, the reader amplitude-modulates the carrier, and the tag simultaneously harvests energy and demodulates the command. During a tag-reply interval, the reader supplies an unmodulated continuous wave that powers and illuminates the tag for backscatter.
A typical interrogation begins when tags become energized in the coverage area of the reader. The reader may use a \textit{Select} command to address a tag subpopulation, followed by an inventory round in which \textit{Query}, \textit{QueryRep}, and \textit{QueryAdjust} commands singulate individual tags. A singulated tag first backscatters an \textit{RN16}; the reader echoes it in an \textit{ACK}, after which the tag returns its EPC identifier.
Successful interrogation requires the tag to remain powered and decode the forward-link commands, and the reader to reliably detect the backscattered reply.

\subsection{Signal Model}
The transmit signal at the access point, denoted by $\bx \in \bbC^{M \times 1}$, is a sum of the sensing signal for tag interrogation and the communication signals for the users, expressed as
\begin{align}
    \bx &= \bff_t^{(\rm{s})} s_t^{(\rm{s})} + \sum_{u=1}^{U} \bff_u^{(\rm{c})} s_u^{(\rm{c})},
\end{align} 
where $s_t^{(\rm{s})} \in \bbC$ represents the RFID interrogation signal for the $t^{\rm{th}}$ tag (i.e., a modulated signal in the command interval and a continuous wave in the tag-reply interval), and $s_u^{(\rm{c})} \in \bbC$ denotes the data symbol for the $u^{\rm{th}}$ communication user. We assume that the interrogation waveform and communication symbols are mutually uncorrelated and normalized to unit average power, such that $\bbE[|s_t^{(\rm{s})}|^2]=\bbE[|s_u^{(\rm{c})}|^2]=1$. The vectors $\bff_t^{(\rm{s})} \in \bbC^{M \times 1}$ and $\bff_u^{(\rm{c})} \in \bbC^{M \times 1}$ are the corresponding beamformers for sensing and communication, respectively. 
The allocated sensing and communication powers are defined as the squared Euclidean norms of their corresponding beamforming vectors, i.e., $P_t^{(\rm{s})} = \|\bff_t^{(\rm{s})}\|^2$ and $P_u^{(\rm{c})} = \|\bff_u^{(\rm{c})}\|^2$.
These beamformers are subject to a total transmit power constraint $P$, given by $P_t^{(\rm{s})} + \sum_{u=1}^{U} P_u^{(\rm{c})} \leq P$. It is worth noting that, although the interrogation waveform alternates between modulated commands and continuous-wave illumination, we assume that the spatial beamformers remain fixed throughout the command–reply transaction.

\subsection{Backscatter Model} \label{subsec:backscatter_model}
The tag receives the interrogation and communication signals in both command and reply intervals. It decodes reader commands in command intervals and modulates the composite incident signal in its assigned reply intervals. We therefore first model the signal received at the tag and then the backscattered signal received at the access point.
Let $\bg_t \in \bbC^{M \times 1}$ denote the channel between the access point and the $t^{\rm{th}}$ tag. The received signal at this tag is expressed as
\begin{align} \label{eq:recv_sig_tag}
    y_t &= \bg_t^H \bx + n_t = \bg_t^H \bff_t^{(\rm{s})} s_t^{(\rm{s})} + \sum_{u=1}^{U} \bg_t^H \bff_u^{(\rm{c})} s_u^{(\rm{c})} + n_t,
\end{align}
where $n_t \sim \cC\cN(0, \sigma_\rm{t}^2)$ represents the incident RF-domain noise at the tag. Subsequently, the tag modulates the impinging signal to scatter back its stored data. The resulting backscatter-modulated signal is $r_t = \sqrt{\eta} \, y_t \, b_t$, where $\eta$ represents the backscatter modulation efficiency, and $b_t$ denotes the encoded tag-data symbol. The encoded tag-data symbol is assumed to be zero-mean, independent of the interrogation signal and communication symbols, and normalized to unit average power. Assuming channel reciprocity between the access point and the tag, the signal received at the access point is
\begin{align} \label{eq:recv_sig_reader}
    y_r &= \bw_t^H \bg_t \, r_t + \bw_t^H \bn_r \nonumber \\
    &= \bw_t^H \bg_t \sqrt{\eta} \, b_t (\bg_t^H \bff_t^{(\rm{s})} s_t^{(\rm{s})} + \sum_{u=1}^{U} \bg_t^H \bff_u^{(\rm{c})} s_u^{(\rm{c})} + n_t) \nonumber \\
	&\quad + \bw_t^H \bn_r,
\end{align}
where $\bw_t \in \bbC^{M \times 1}$ is the unit-norm combining vector used by the access point, and $\bn_r \sim \cC\cN(\bm{0}, \sigma_\rm{r}^2 \bI)$ is the receiver noise vector.
Each tag reply begins with a known preamble \cite{EPCGlobal} that can be used for channel estimation, and we assume that the access point accurately estimates the tag-to-reader channel using the known preamble and applies maximum-ratio combining, i.e., $\bw_t = \bg_t / \|\bg_t\|$. This online channel estimate is used for receive combining and subsequent detection and decoding of the tag reply.
Based on the model in \eqref{eq:recv_sig_tag}, the tag-side SINR for receiving the interrogation signal is given by
\begin{align} \label{eq:SINR_tag}
    &\rm{SINR}_{t}^{(\rm{t})}(\bff_t^{(\rm{s})}, \{\bff_u^{(\rm{c})}\}_{u=1}^{U}) \nonumber \\
	&= \frac{\bbE\left[|\bg_t^H \bff_t^{(\rm{s})} s_t^{(\rm{s})}|^2\right]}{\bbE\left[\sum_{u=1}^{U}|\bg_t^H \bff_u^{(\rm{c})} s_u^{(\rm{c})}|^2\right] + \bbE\left[|n_t|^2\right]} \nonumber \\
    &= \frac{|\bg_t^H \bff_t^{(\rm{s})}|^2}{\sum_{u=1}^{U}|\bg_t^H \bff_u^{(\rm{c})}|^2 + \sigma_\rm{t}^2} \nonumber \\
    &= \frac{P_t^{(\rm{s})}|\bg_t^H \bar{\bff}_t^{(\rm{s})}|^2}{\sum_{u=1}^{U}P_u^{(\rm{c})}|\bg_t^H \bar{\bff}_u^{(\rm{c})}|^2 + \sigma_\rm{t}^2},
\end{align}
where the terms $\bar{\bff}_t^{(\rm{s})}$ and $\bar{\bff}_u^{(\rm{c})}$ are the normalized sensing and communication beamforming vectors, i.e., $\bar{\bff}_t^{(\rm{s})}=\bff_t^{(\rm{s})}/\|\bff_t^{(\rm{s})}\|$, $\bar{\bff}_u^{(\rm{c})}=\bff_u^{(\rm{c})}/\|\bff_u^{(\rm{c})}\|$. 
Similarly, using \eqref{eq:recv_sig_reader}, the SINR for detecting the backscattered tag reply at the access point is given by
\begin{align} \label{eq:SINR_reader}
    &\rm{SINR}_{t}^{(\rm{r})}(\bff_t^{(\rm{s})}, \{\bff_u^{(\rm{c})}\}_{u=1}^{U}) \nonumber \\
	&= \frac{\eta |\bw_t^H \bg_t|^2 |\bg_t^H \bff_t^{(\rm{s})}|^2}{\sum_{u=1}^{U} \eta |\bw_t^H \bg_t|^2 |\bg_t^H \bff_u^{(\rm{c})}|^2 + \eta \, \sigma_\rm{t}^2 |\bw_t^H \bg_t|^2 + \sigma_\rm{r}^2} \nonumber \\
    &= \frac{\eta |\bw_t^H \bg_t|^2 P_t^{(\rm{s})} |\bg_t^H \bar{\bff}_t^{(\rm{s})}|^2}{\sum_{u=1}^{U} \eta |\bw_t^H \bg_t|^2 P_u^{(\rm{c})} |\bg_t^H \bar{\bff}_u^{(\rm{c})}|^2 + \eta \, \sigma_\rm{t}^2 |\bw_t^H \bg_t|^2 + \sigma_\rm{r}^2}.
\end{align}
In the considered ISAC system, the communication signals have two effects on RFID interrogation. First, their incident RF power can contribute to powering the passive tag during both the command and tag-reply intervals. Second, because these signals do not carry the RFID interrogation waveform, they can impair command detection at the tag during the command interval. During the tag-reply interval, the tag backscatters the composite incident waveform; therefore, the backscattered components originating from the communication signals also affect detection of the tag reply at the access point and are modeled as interference in \eqref{eq:SINR_reader}.

Commercial RFID tag sensitivity is typically specified in terms of the minimum received power~\cite{Impinj_Tag}, and the minimum SINR required for a tag to reliably decode reader commands has not been comprehensively characterized and may vary with the interference conditions~\cite{Krigslund2011}. Therefore, we use the available tag power sensitivity specification to set $\gamma_\rm{t}=P_{\rm{sen,t}}/\sigma_\rm{t}^2$ as a conservative sufficient condition, where $P_{\rm{sen,t}}$ is the received power sensitivity of the tag. 
With this threshold, the tag-side condition $\rm{SINR}_{t}^{(\rm{t})}(\bff_t^{(\rm{s})}, \{\bff_u^{(\rm{c})}\}_{u=1}^{U}) \geq \gamma_\rm{t}$ is equivalent to
\begin{align}
    |\bg_t^H \bff_t^{(\rm{s})}|^2 \geq P_{\rm{sen,t}} \bigg(1 + \frac{\sum_{u=1}^{U}|\bg_t^H \bff_u^{(\rm{c})}|^2}{\sigma_\rm{t}^2}\bigg).
\end{align}
In the absence of communication interference, this condition reduces exactly to the tag power sensitivity requirement. When communication signals are present, it requires an additional margin on the interrogation signal power that grows with the ratio of the communication interference power to the tag noise power. Hence, the condition guarantees that the tag is powered by the interrogation signal alone and that the communication interference at the tag is weak compared with the interrogation signal.
Although the communication signals may also supply incident RF energy, the adopted condition does not credit this energy toward tag powering. Consequently, the condition is sufficient but not necessary for satisfying the tag power sensitivity requirement and provides a conservative, interference-aware surrogate for tag-side operation.

During the tag-reply interval, the access point must reliably detect the backscattered tag signal in the presence of the backscattered components induced by the communication signals. We adopt an analogous condition at the reader by setting $\gamma_\rm{r}=P_{\rm{sen,r}}/\sigma_\rm{r}^2$, where $P_{\rm{sen,r}}$ is the reader sensitivity~\cite{Impinj_Reader}. In the absence of communication interference, satisfying $\rm{SINR}_{t}^{(\rm{r})}(\bff_t^{(\rm{s})}, \{\bff_u^{(\rm{c})}\}_{u=1}^{U}) \geq \gamma_\rm{r}$ guarantees that the backscattered signal reaches the reader-sensitivity level. When communication signals are present, the same condition imposes an additional margin against their backscattered interference and is also a design surrogate. A tag is successfully interrogated only when both tag-side and reader-side conditions are satisfied using the same beamformers during the corresponding command-reply transaction.

\subsection{Communication Model}
The communication users are served throughout the RFID command-reply transaction. In both the command and tag-reply intervals, a communication user experiences direct interference from the RFID interrogation signal and the other user data streams. In the tag-reply interval, the replying tag also backscatters the composite incident waveform, introducing additional interference. Since the spatial beamformers remain fixed throughout the transaction, the tag-reply interval is the worse of the two intervals in terms of communication user SINR. We therefore formulate the user SINR for the tag-reply interval, whose requirement also guarantees that of the command interval.
Accordingly, during the tag-reply interval, the $u^{\rm{th}}$ communication user receives a superposition of the signals transmitted directly by the access point and the signal backscattered by the tag. The received signal is expressed as
\begin{align}
    y_u &= \bh_u^H \bx + h_{t,u} r_t + n_u \nonumber \\
    &= \bh_u^H \bff_u^{(\rm{c})} s_u^{(\rm{c})} +  \sum_{l=1, l \neq u}^{U} \bh_u^H \bff_l^{(\rm{c})} s_l^{(\rm{c})} + \bh_u^H \bff_t^{(\rm{s})} s_t^{(\rm{s})} \nonumber \\
    & \quad + h_{t,u} \sqrt{\eta} \, b_t(\bg_t^H \bff_t^{(\rm{s})} s_t^{(\rm{s})} + \sum_{v=1}^{U} \bg_t^H \bff_v^{(\rm{c})} s_v^{(\rm{c})} + n_t) + n_u,
\end{align}
where $\bh_u \in \bbC^{M \times 1}$ denotes the channel between the access point and the $u^{\rm{th}}$ user, $h_{t,u} \in \bbC$ represents the channel between the $t^{\rm{th}}$ tag and the $u^{\rm{th}}$ user, and $n_u \sim \cC\cN(0, \sigma_\rm{u}^2)$ is the receiver noise at the user. 
Accordingly, the communication user SINR for decoding the intended downlink data stream is given in \eqref{eq:SINR_user}.
For notational brevity, the beamforming arguments of the SINR functions are omitted in subsequent formulations when they are clear from the context.
\begin{figure*}[!t]
    \begin{align} \label{eq:SINR_user}
        &\rm{SINR}_{t,u}^{(\rm{c})}(\bff_t^{(\rm{s})}, \{\bff_u^{(\rm{c})}\}_{u=1}^{U}) \nonumber \\
		&= \frac{|\bh_u^H \bff_u^{(\rm{c})}|^2}{\sum_{l \neq u}^{U} |\bh_u^H \bff_l^{(\rm{c})}|^2 + |\bh_u^H \bff_t^{(\rm{s})}|^2 + \eta |h_{t,u}|^2 (|\bg_t^H \bff_t^{(\rm{s})}|^2 + \sum_{v=1}^{U} |\bg_t^H \bff_v^{(\rm{c})}|^2 + \sigma_\rm{t}^2) + \sigma_\rm{u}^2} \nonumber \\
        &= \frac{P_u^{(\rm{c})} |\bh_u^H \bar{\bff}_u^{(\rm{c})}|^2}{\sum_{l \neq u}^{U} P_l^{(\rm{c})} |\bh_u^H \bar{\bff}_l^{(\rm{c})}|^2 + P_t^{(\rm{s})} |\bh_u^H \bar{\bff}_t^{(\rm{s})}|^2 + \eta |h_{t,u}|^2 (P_t^{(\rm{s})} |\bg_t^H \bar{\bff}_t^{(\rm{s})}|^2 + \sum_{v=1}^{U} P_v^{(\rm{c})} |\bg_t^H \bar{\bff}_v^{(\rm{c})}|^2 + \sigma_\rm{t}^2) + \sigma_\rm{u}^2}.
    \end{align}
	\hrule
\end{figure*}

\section{Problem Formulation} \label{sec:problem_formulation}
We aim to design a joint sensing and communication beamforming codebook. During beam sweeping, the access point sequentially transmits a set of codewords, each consisting of one sensing beam and one communication beam for each user. Let $\cF=\{(\bff_n^{(\rm{s})}, \{\bff_{n,u}^{(\rm{c})}\}_{u=1}^{U})\}_{n=1}^{N}$ define the joint codebook with cardinality $|\cF|=N$, and $\mathcal{N} = \{1, 2, \ldots, N\}$ denote the set of all possible codebook indices.
Consequently, the codebook design problem can be formulated as 
\begin{subequations} \label{eq:opt}
\begin{align}
    \max_{\cF}\quad
    & \sum_{t=1}^{T} I_t
    \\
    \textrm{s.t.}\quad
    & I_t =
      \max_{n\in\mathcal{N}}
      \left\{
      \mathds{1}\left(
          \rm{SINR}_{t,n}^{(\rm{t})}
          \geq \gamma_\rm{t}
      \right)
      \mathds{1}\left(
          \rm{SINR}_{t,n}^{(\rm{r})}
          \geq \gamma_\rm{r}
      \right)
      \right\},
    \nonumber
    \\
	& \forall t \in \mathcal{T}, \label{eq:opt_constraint_1} \\
    & \rm{SINR}_{t,u,n}^{(\rm{c})}
      \geq \gamma_\rm{u},
      \quad \forall t \in \mathcal{T}, u \in \mathcal{U}, n \in \mathcal{N},
      \label{eq:opt_constraint_2}
    \\
    & \|\bff_n^{(\rm{s})}\|^2
      + \sum_{u=1}^{U}
        \|\bff_{n,u}^{(\rm{c})}\|^2
      \leq P,
      \quad \forall n \in \mathcal{N}.
      \label{eq:opt_constraint_3}
\end{align}
\end{subequations}
Here, the interrogation indicator for the $t^{\rm{th}}$ tag, denoted by $I_t$, equals one if there exists at least one codeword that simultaneously satisfies the tag-side and reader-side SINR requirements, and equals zero otherwise. For compactness, the subscript $n$ on an SINR term denotes its evaluation using the $n^{\rm{th}}$ codeword. The objective therefore maximizes the number of successfully interrogated tags over the complete beam sweep. The first constraint \eqref{eq:opt_constraint_1} ensures that the two SINR requirements are met by the same codeword, where $\mathds{1}(\cdot)$ denotes the indicator function.
Constraint \eqref{eq:opt_constraint_2} guarantees that the communication SINR requirement of every user is satisfied for each codeword, while constraint \eqref{eq:opt_constraint_3} imposes the total transmit power limit for each codeword. At the time of codebook design, we assume that the access point has knowledge of the user channels $\{\bh_u\}_{u=1}^{U}$, which can be acquired through standard pilot-based channel estimation during the offline stage.

The codebook design problem, however, presents two main challenges. First, the interrogation and communication transmissions are coupled through mutual interference. The communication signals interfere with command decoding at the tag and reply detection at the reader, whereas the interrogation and backscattered signals can degrade communication reception. Consequently, the sensing and communication beams must be jointly designed to balance the interrogation and communication requirements. Second, the positions and channel information of the actual tags are generally unavailable, so the tag-dependent quantities in \eqref{eq:opt} cannot be evaluated directly. To address these challenges, \sref{sec:codebook_design_multi_tag} proposes a site-specific codebook design that recasts the problem as a sector-wise spatial coverage problem and employs grid sampling for tractability. Before that, \sref{sec:beamforming_design_single_tag} studies the single-tag beamforming design with known channel information, which provides insights into mutual interference and interrogation feasibility and lays the foundation for the codebook design.

\section{Beamforming Design for the Single-Tag Scenario} \label{sec:beamforming_design_single_tag}
Before delving into the codebook design problem, we first explore the fundamental problem of beamforming for a single tag, i.e., $T=1$. In this scenario, we assume the access point has knowledge of the tag channel $\bg_t$ for beamforming design. The objective is to design a transmit sensing beam $\bff_t^{(\rm{s})}$ and communication beams $\{\bff_u^{(\rm{c})}\}_{u=1}^{U}$ that can successfully interrogate the tag while satisfying the SINR requirements of the communication users. Thus, the optimization problem in \eqref{eq:opt} reduces to the following feasibility check problem:
\begin{gather} \label{eq:single_tag_feas}
	\begin{aligned}
		\rm{find} \quad & (\bff_t^{(\rm{s})}, \{\bff_u^{(\rm{c})}\}_{u=1}^{U}) \\
		\textrm{s.t.} \quad & \rm{SINR}_t^{(\rm{t})} \geq \gamma_\rm{t}, \\
		\quad & \rm{SINR}_t^{(\rm{r})} \geq \gamma_\rm{r},  \\
		\quad & \rm{SINR}_{t,u}^{(\rm{c})} \geq \gamma_\rm{u}, \ \forall u \in \mathcal{U}, \\
	\quad & \|\bff_t^{(\rm{s})}\|^2 + \sum_{u=1}^{U} \|\bff_u^{(\rm{c})}\|^2 \leq P.
	\end{aligned}
\end{gather} 
To address this problem, we propose two distinct beamforming strategies. First, we introduce a zero-forcing approach combined with power allocation optimization, for which we derive the dominant constraint and a closed-form solution. Second, we present a general joint design of sensing and communication beams through transmit power minimization.

\subsection{Solution 1: Zero-Forcing Beamforming and Power Allocation} \label{subsec:ZF_beamforming}
In this subsection, we propose a two-stage approach: first designing the beamforming vectors using zero-forcing~\cite{Bjornson2014}, and then optimizing the power allocation between sensing and communication. To eliminate mutual interference, we project the intended signal for each stream onto the null space of the others. Let $\bH = [\bg_t, \bh_1, \ldots, \bh_U] \in \bbC^{M \times (U+1)}$ denote the aggregate channel matrix containing both the tag and user channels. 
We assume that the aggregate channel matrix has full column rank, which requires the number of access-point antennas to be at least the number of communication users plus one. Thus, the access point has sufficient spatial degrees of freedom to construct the sensing and communication beams satisfying the zero-forcing nulling conditions.
Under this assumption, the zero-forcing beamforming vectors are computed via the pseudo-inverse of $\bH$, given by
\begin{equation}
    [\tilde{\bff}_t^{(\rm{s})}, \tilde{\bff}_1^{(\rm{c})}, \ldots, \tilde{\bff}_U^{(\rm{c})}] = \bH (\bH^H \bH)^{-1}.
\end{equation}
These vectors are then normalized and scaled by the allocated power, such that $\bff_t^{(\rm{s})} = \sqrt{P_t^{(\rm{s})}} \, (\tilde{\bff}_t^{(\rm{s})} / \|\tilde{\bff}_t^{(\rm{s})}\|), \bff_u^{(\rm{c})} = \sqrt{P_u^{(\rm{c})}} \, (\tilde{\bff}_u^{(\rm{c})} / \|\tilde{\bff}_u^{(\rm{c})}\|), \forall u \in \mathcal{U}$. Since the feasibility problem in \eqref{eq:single_tag_feas} may yield multiple solutions, we aim to find the most power-efficient one. Accordingly, we reformulate the problem as a power minimization problem:
\begin{subequations} \label{eq:opt_PA}
	\begin{align}
		\min_{P_t^{(\rm{s})}, \{P_u^{(\rm{c})}\}_{u=1}^{U}} \quad & P_t^{(\rm{s})} + \sum_{u=1}^{U} P_u^{(\rm{c})} \\
		\textrm{s.t.} \quad & \rm{SINR}_t^{(\rm{t})} \geq \gamma_\rm{t}, \label{eq:opt_PA_constraint_1} \\
		\quad & \rm{SINR}_t^{(\rm{r})} \geq \gamma_\rm{r},  \label{eq:opt_PA_constraint_2} \\
		\quad & \rm{SINR}_{t,u}^{(\rm{c})} \geq \gamma_\rm{u}, \ \forall u \in \mathcal{U}, \label{eq:opt_PA_constraint_3} \\
        \quad & P_t^{(\rm{s})} + \sum_{u=1}^{U} P_u^{(\rm{c})} \leq P. \label{eq:opt_PA_constraint_4}
	\end{align}
\end{subequations}
After fixing these beamforming directions using zero-forcing, the remaining SINR constraints are linear in the sensing and communication powers, and the power allocation problem is a linear program. Also, every communication beam is nulled at the tag and the unintended users, while the sensing beam is nulled at all communication users. Consequently, the direct inter-beam interference terms vanish. This structure simplifies the analysis, allowing us to better understand the problem and derive a closed-form solution. We begin by establishing the following proposition to identify the dominant constraint for tag interrogation.

\noindent \textbf{Proposition 1.} \textit{Under the zero-forcing design described above, the minimum sensing power required to satisfy both RFID interrogation constraints is determined by the larger of the tag-side and reader-side SINR requirements. Specifically, the required interrogation power is given by}
	\begin{equation} 
		P_t^{(\rm{s})} \geq
		\begin{cases}
			\frac{\gamma_\rm{t} \sigma_\rm{t}^2}{|\bg_t^H \bar{\bff}_t^{(\rm{s})}|^2} & \text{if \eqref{eq:opt_PA_constraint_1} is dominant},\\
			\frac{\gamma_\rm{r} (\eta \, \sigma_\rm{t}^2 |\bw_t^H \bg_t|^2 + \sigma_\rm{r}^2)}{\eta |\bw_t^H \bg_t|^2 |\bg_t^H \bar{\bff}_t^{(\rm{s})}|^2} & \text{otherwise}.
		\end{cases}
	\end{equation}
\textit{The tag-side constraint is dominant when its power lower bound is no smaller than the reader-side lower bound; otherwise, the reader-side constraint is dominant. Therefore, selecting the larger of the two bounds guarantees that both SINR constraints are satisfied.}

\noindent The proof is detailed in Appendix \ref{appendix:proof_proposition_1}. 
Proposition 1 characterizes the minimum sensing power required to satisfy the two RFID constraints under the zero-forcing design. It does not by itself guarantee feasibility of the complete power allocation problem, which additionally requires satisfaction of the communication user SINR constraints and the total transmit power constraint. If the required sensing power alone exceeds the total power budget, the zero-forcing power allocation problem is necessarily infeasible.
As will be demonstrated in the evaluation section, the tag-side SINR constraint is often dominant, highlighting the critical role of transmit beamforming design in enhancing the power delivered to the tag. 
Another important observation is that, under the zero-forcing design, the minimum-power solution must satisfy the dominant RFID SINR constraint and all communication user SINR constraints with equality.
Building on this insight, we proceed to derive a closed-form solution for the power allocation problem in \eqref{eq:opt_PA}, assuming one of the constraints is dominant.

\noindent \textbf{Proposition 2.} \textit{Assuming the tag-side SINR constraint \eqref{eq:opt_PA_constraint_1} is dominant for the interrogation, the optimal power allocation that minimizes the total transmit power can be expressed as
	\begin{subequations} \label{eq:opt_PA_solution}
		\begin{align} 
			P_t^{(\rm{s})} &= \frac{\gamma_\rm{t} \sigma_\rm{t}^2}{|\bg_t^H \bar{\bff}_t^{(\rm{s})}|^2}, \label{eq:opt_PA_solution_sensing} \\
			P_u^{(\rm{c})} &= \frac{\gamma_\rm{u} \left[ \eta |h_{t,u}|^2 \sigma_\rm{t}^2 (\gamma_\rm{t} + 1) + \sigma_\rm{u}^2 \right]}{|\bh_u^H \bar{\bff}_u^{(\rm{c})}|^2}, \ \forall u \in \mathcal{U}. \label{eq:opt_PA_solution_comm}
		\end{align}
	\end{subequations}
\textit{The resulting allocation is feasible and optimal for the fixed zero-forcing beams if the sum of the required sensing and communication powers does not exceed the total power budget. If this sum exceeds the power budget, the zero-forcing power-allocation problem is infeasible.}}

\noindent The proof is provided in Appendix \ref{appendix:proof_proposition_2}.
This proposition offers a straightforward method to compute the optimal power allocation for sensing and communication under the zero-forcing beamforming framework. Notably, the optimal power allocation aims to compensate for the channel gains and meet the SINR requirements, given the established beamforming directions. Also, as we will see in the simulation results, this zero-forcing-based approach achieves performance close to that of the optimal joint design, particularly when the communication SINR requirements are high.

\subsection{Solution 2: Joint Beamforming Design via Transmit Power Minimization} \label{subsec:joint_beamforming}
In this subsection, we generalize the design by jointly optimizing the beamforming vectors and power allocation for both sensing and communication. By directly incorporating the beamforming vectors into the framework of \eqref{eq:opt_PA}, we formulate the joint beamforming optimization problem as follows:
\begin{subequations} \label{eq:single_tag_opt_BF}
	\begin{align}
		\min_{\bff_t^{(\rm{s})}, \{\bff_u^{(\rm{c})}\}_{u=1}^{U}} \quad & \|\bff_t^{(\rm{s})}\|^2 + \sum_{u=1}^{U} \|\bff_u^{(\rm{c})}\|^2 \\
		\textrm{s.t.} \quad & \rm{SINR}_t^{(\rm{t})} \geq \gamma_\rm{t}, \label{eq:single_tag_opt_BF_constraint_1} \\
		\quad & \rm{SINR}_t^{(\rm{r})} \geq \gamma_\rm{r}, \label{eq:single_tag_opt_BF_constraint_2} \\
		\quad & \rm{SINR}_{t,u}^{(\rm{c})} \geq \gamma_\rm{u}, \ \forall u \in \mathcal{U}, \label{eq:single_tag_opt_BF_constraint_3} \\
        \quad & \|\bff_t^{(\rm{s})}\|^2 + \sum_{u=1}^{U} \|\bff_u^{(\rm{c})}\|^2 \leq P.
	\end{align}
\end{subequations}
Due to the fractional forms of SINR constraints,  \eqref{eq:single_tag_opt_BF} is non-convex. However, these constraints can be cast into second-order cone constraints~\cite{Demirhan2023b}. Taking the user SINR constraint \eqref{eq:single_tag_opt_BF_constraint_3} as an example, we first rearrange the terms as follows:
\begin{align} \label{eq:single_tag_opt_BF_constraint_3_rewritten}
    &\frac{1}{\gamma_\rm{u}} |\bh_u^H \bff_u^{(\rm{c})}|^2 \geq \sum_{l \neq u}^{U} |\bh_u^H \bff_l^{(\rm{c})}|^2 + |\bh_u^H \bff_t^{(\rm{s})}|^2 \\
	&\quad+ \eta |h_{t,u}|^2 (|\bg_t^H \bff_t^{(\rm{s})}|^2 + \sum_{v=1}^{U} |\bg_t^H \bff_v^{(\rm{c})}|^2 + \sigma_\rm{t}^2) + \sigma_\rm{u}^2. \nonumber
\end{align}
The non-linearity of the absolute value term on the left-hand side poses a challenge. To resolve this, we leverage the property that an arbitrary phase rotation can be applied to the expression within the absolute value without changing its magnitude. Specifically, we have $|\bh_u^H \bff_u^{(\rm{c})}| = |\bh_u^H \bff_u^{(\rm{c})} \E^{j \theta}|$. Without loss of optimality, we apply a phase rotation $\E^{j \theta}$ such that $\bh_u^H \bff_u^{(\rm{c})}$ becomes real and non-negative, i.e., $|\bh_u^H \bff_u^{(\rm{c})}| = \RE\left\{\bh_u^H \bff_u^{(\rm{c})}\right\}$.
This operation effectively couples the search for the optimal beamforming vector with the associated phase rotation. Consequently, taking the square root of both sides in \eqref{eq:single_tag_opt_BF_constraint_3_rewritten} yields the following second-order cone constraint:
\begin{align} \label{eq:reformulated_SINR_u}
    & \sqrt{\frac{1}{\gamma_\rm{u}}} \textrm{Re}\left\{ \bh_u^H \bff_u^{(\rm{c})} \right\} \geq \bigg( \sum_{l \neq u}^{U} |\bh_u^H \bff_l^{(\rm{c})}|^2 + |\bh_u^H \bff_t^{(\rm{s})}|^2  \\
	&\quad+ \eta |h_{t,u}|^2 (|\bg_t^H \bff_t^{(\rm{s})}|^2 + \sum_{v=1}^{U} |\bg_t^H \bff_v^{(\rm{c})}|^2 + \sigma_\rm{t}^2) + \sigma_\rm{u}^2 \bigg)^{\frac{1}{2}}. \nonumber
\end{align}
Applying similar transformations to the tag-side and reader-side SINR constraints in \eqref{eq:single_tag_opt_BF_constraint_1} and \eqref{eq:single_tag_opt_BF_constraint_2}, we can obtain the following second-order cone constraints:
\begin{equation} \label{eq:reformulated_SINR_t}
    \sqrt{\frac{1}{\gamma_\rm{t}}} \textrm{Re}\left\{ \bg_t^H \bff_t^{(\rm{s})} \right\} \geq
    \begin{Vmatrix}
        \bg_t^H \bff_1^{(\rm{c})}\\
		\vdots \\
		\bg_t^H \bff_U^{(\rm{c})}\\
        \sigma_\rm{t}
    \end{Vmatrix},
\end{equation}

\begin{equation} \label{eq:reformulated_SINR_r}
    \sqrt{\frac{\eta}{\gamma_\rm{r}}} |\bw_t^H \bg_t| \textrm{Re}\left\{ \bg_t^H \bff_t^{(\rm{s})} \right\} \geq
    \begin{Vmatrix}
        \sqrt{\eta} \bw_t^H \bg_t \bg_t^H \bff_1^{(\rm{c})}\\
		\vdots \\ 
		\sqrt{\eta} \bw_t^H \bg_t \bg_t^H \bff_U^{(\rm{c})}\\
        \sqrt{\eta} \bw_t^H \bg_t \sigma_\rm{t} \\
        \sigma_\rm{r}
    \end{Vmatrix}.
\end{equation}
The reformulated constraints in \eqref{eq:reformulated_SINR_u}-\eqref{eq:reformulated_SINR_r} are mathematically equivalent to the original constraints. Although the tag-side and reader-side constraints contain different interference and noise terms, their desired signal terms depend on the same inner product between the tag channel and the sensing beamforming vector. 
Moreover, the objective and all constraints are invariant to a common phase rotation of the sensing beam. We can therefore select its phase such that this common inner product is real and nonnegative. The phase rotation of each communication beam can be independently selected with respect to its intended user without affecting the other constraints. Consequently, all three types of SINR constraints admit second-order-cone representations, and the resulting joint beamforming optimization problem is therefore convex and can be solved as a second-order-cone program.

\section{Beamforming Codebook Design for the Multi-Tag Scenario} \label{sec:codebook_design_multi_tag}
Having established the beamforming design for the single-tag scenario, we now present the site-specific codebook design for the multi-tag scenario. To obtain a tractable solution to the codebook design problem in \eqref{eq:opt}, we split the region of interest into sectors and design one sensing and communication codeword for each sector. The interrogation coverage optimization problem for each sector is defined using a finite grid of reference locations, and the sector-specific codewords are finally combined to form the complete codebook.

\subsection{Site-Specific Beamforming Codebook Design}
We first formulate the beamforming design problem for a generic
sector. We model the region of interest as a semi-annular area centered at the access point, with inner and outer radial boundaries $R_{\rm{min}}$ and $R_{\rm{max}}$, respectively. These boundaries specify the spatial region over which the sector codeword is designed.
Each sector is defined by an angular interval $[\Theta_{\rm{min}},\Theta_{\rm{max}}]$ and the radial interval $[R_{\rm{min}},R_{\rm{max}}]$.
Let $\mathcal{S} = \left\{(r,\theta):R_\rm{min}\leq r\leq R_\rm{max},\, \Theta_{\rm{min}}\leq\theta\leq\Theta_{\rm{max}}\right\}$ denote the set of potential tag locations within the sector. For each potential tag location $(r,\theta)\in\mathcal{S}$, we define the tag-side and reader-side binary indicators, respectively, as
\begin{gather}
	\begin{aligned}
		y_{r,\theta}^{(\rm{t})}
		&=
		\mathds{1}\left(
			\rm{SINR}_{r,\theta}^{(\rm{t})}
			\geq \gamma_\rm{t}
		\right),
		\\
		y_{r,\theta}^{(\rm{r})}
		&=
		\mathds{1}\left(
			\rm{SINR}_{r,\theta}^{(\rm{r})}
			\geq \gamma_\rm{r}
		\right).
	\end{aligned}
\end{gather}
Accordingly, $y_{r,\theta}^{(\rm{t})} \, y_{r,\theta}^{(\rm{r})}=1$ if and only if both RFID SINR requirements are satisfied at $(r,\theta)$ under the same sensing and communication beams. The beamforming design problem for each sector is then formulated as follows.
\begin{IEEEeqnarray}{cl} \label{eq:opt_BF}
	\max_{\bff^{(\rm{s})}, \{\bff_u^{(\rm{c})}\}_{u=1}^{U}} \quad & \int_{\Theta_\rm{min}}^{\Theta_\rm{max}} \int_{R_\rm{min}}^{R_\rm{max}}  y_{r, \theta}^{(\rm{t})} \, y_{r, \theta}^{(\rm{r})} \, r\,dr\,d\theta \nonumber \\
	\textrm{s.t.} \quad & \rm{SINR}_{r,\theta,u}^{(\rm{c})} \geq\gamma_\rm{u}, \, \forall (r,\theta)\in\mathcal{S}, \, \forall u \in \mathcal{U}, \nonumber \\
	\quad & \|\bff^{(\rm{s})}\|^2 + \sum_{u=1}^{U} \|\bff_u^{(\rm{c})}\|^2 \leq P.
\end{IEEEeqnarray}
Here, $\rm{SINR}_{r,\theta,u}^{(\rm{c})}$ denotes the SINR of the $u^{\rm{th}}$ communication user in \eqref{eq:SINR_user} when the replying tag is located at $(r,\theta)$, and the first constraint ensures that the communication user SINR requirement is satisfied for every potential tag location in $\mathcal{S}$. All SINRs in \eqref{eq:opt_BF} are evaluated using the sector-specific sensing and communication beams optimized in that problem. The objective maximizes the interrogation coverage within the sector while satisfying the communication user and transmit power requirements. This sector-wise formulation provides a tractable approximation of the codebook design problem in \eqref{eq:opt} in two ways. First, the objective is expressed in terms of the interrogation coverage, rather than the number of actual tags. This allows us to design a codeword that maximizes the potential interrogation coverage without requiring knowledge of the actual tag locations. Second, the spatial domain is reduced to a single sector, which is then solved independently of other sectors. The connection between this sector-wise design and the codebook design in \eqref{eq:opt} is presented in Appendix \ref{appendix:connection_opt_sector}. In particular, when the tags are uniformly distributed over the region of interest, the objective of \eqref{eq:opt} in expectation can be expressed in terms of the spatial coverage, i.e., the area covered by at least one codeword, and this spatial coverage is lower bounded by the sum of the sector-wise objectives in \eqref{eq:opt_BF} over all sectors.

The continuous spatial domain and indicator functions in \eqref{eq:opt_BF}, however, make the problem difficult to solve directly. To obtain a finite-dimensional approximation, we sample the sector at a grid of reference locations $\mathcal{G}$ with $K_r\geq 1$ radial and $K_\theta\geq 2$ angular samples. 
Since the area element in \eqref{eq:opt_BF} is $r\,dr\,d\theta$, the area of an annular sector between radii $r_1$ and $r_2$ is proportional to $r_2^2-r_1^2$. We therefore space the radial samples uniformly in $r^2$, which divides the sector into $K_r$ annular sectors of equal area, each represented by one radial sample. As a result, counting the covered reference locations approximates the covered area in \eqref{eq:opt_BF}. The angular domain is sampled uniformly, including its boundary points. 
The corresponding sampling intervals are defined as
\begin{equation}
    \Delta_r = \frac{R_{\max}^2-R_{\min}^2}{K_r},
    \qquad
    \Delta_\theta = \frac{\Theta_{\max}-\Theta_{\min}}{K_\theta-1}.
\end{equation}
The resulting radial sampling set $\mathcal{R}$, angular sampling set
$\vartheta$, and the grid $\mathcal{G}$ are defined as
\begin{gather}
	\begin{aligned}
		\mathcal{R} &=
		\left\{
		\sqrt{R_{\min}^2+\left(m-\tfrac{1}{2}\right)\Delta_r}:
		m\in\{1,2,\ldots,K_r\}
		\right\},
		\\
		\vartheta &=
		\left\{
		\Theta_{\min}+n\Delta_\theta:
		n\in\{0,1,\ldots,K_\theta-1\}
		\right\},
		\\
		\mathcal{G} &=
		\mathcal{R}\times\vartheta.
	\end{aligned}
\end{gather}
The resulting grid contains $|\mathcal{G}|=K_rK_\theta$ reference locations within the sector, indexed by $i\in\mathcal{I}=\{1,2,\ldots,|\mathcal{G}|\}$.
Evaluating the SINR conditions at the sampled reference locations requires the corresponding access-point-to-tag channels and tag-to-user channels. We assume that the channel information at the reference locations is obtained during an offline channel acquisition stage.
Temporary reference tags are placed at the selected grid locations and interrogated. At each location, the backscatter observations received at the access point are used to estimate the access-point-to-tag channel $\mathbf{g}_i$. Meanwhile, the communication users measure the corresponding backscattered signals and report their observations to the access point, enabling estimation of the tag-to-user channels $\{h_{i,u}\}_{u=1}^{U}$. Collecting these location-indexed channel measurements over the region of interest forms the offline channel database $\mathcal{D}$. For the sector under consideration, the relevant subset of the database is defined as $\mathcal{D}_{\mathcal{S}} = \left\{\left((r_i,\theta_i),\mathbf{g}_i,\{h_{i,u}\}_{u=1}^{U}\right)\right\}_{i=1}^{|\mathcal{G}|}$. The reference locations serve only as spatial samples for offline codebook design, and the resulting codebook is intended to provide continuous spatial coverage. Therefore, actual tags are not constrained to the reference grid and can be interrogated at arbitrary locations within the region of interest.

Using the channel records in $\mathcal{D}_{\mathcal{S}}$, we introduce a binary indicator vector $\by\in\bbR^{|\mathcal{G}|\times 1}$, where $[\by]_i=1$ indicates that the sensing and communication beams satisfy both the tag-side and reader-side SINR requirements when evaluated using the $i^{\rm{th}}$ record. Our objective is to maximize the number of successfully interrogated reference locations, thereby obtaining a sampling-based surrogate for the interrogation coverage within the sector. Accordingly, the continuous problem in \eqref{eq:opt_BF} is approximated by the following discrete problem.
\begin{IEEEeqnarray}{cl} \label{eq:opt_BF_MINLP}
		\max_{\bff^{(\rm{s})}, \{\bff_u^{(\rm{c})}\}_{u=1}^{U}, \by} \quad &  \sum_{i=1}^{|\cG|} \, [\by]_i \nonumber \\
		\textrm{s.t.} \quad & \rm{SINR}_{i}^{(\rm{t})} \geq \gamma_\rm{t} [\by]_i, \ \forall i \in \mathcal{I}, \nonumber \\
		\quad & \rm{SINR}_{i}^{(\rm{r})} \geq \gamma_\rm{r} [\by]_i, \ \forall i \in \mathcal{I}, \nonumber \\
		\quad & \rm{SINR}_{i,u}^{(\rm{c})} \geq \gamma_\rm{u}, \ \forall i \in \mathcal{I}, u \in \mathcal{U}, \nonumber \\
		\quad & \|\bff^{(\rm{s})}\|^2 + \sum_{u=1}^{U} \|\bff_u^{(\rm{c})}\|^2 \leq P, \nonumber \\
		\quad & [\by]_{i} \in \{0, 1\}, \ \forall i \in \mathcal{I}.
\end{IEEEeqnarray}
Here, the subscript $i$ indicates that the corresponding SINR is evaluated using the channel record associated with the $i^{\rm{th}}$ reference location $(r_i,\theta_i)\in\mathcal{G}$. For instance, $\rm{SINR}_{i}^{(\rm{t})}$ denotes the tag-side SINR at that location.
Due to (i) the mixture of binary variables $\by$ and continuous variables $\bff^{(\rm{s})}$, $\{\bff_u^{(\rm{c})}\}_{u=1}^{U}$ and (ii) the second-order terms in the SINR expressions, \eqref{eq:opt_BF_MINLP} is a mixed-integer nonlinear programming (MINLP) problem~\cite{Floudas1995}. 

Similar to \eqref{eq:single_tag_opt_BF}, the SINR expressions in \eqref{eq:opt_BF_MINLP} are non-convex. However, we cannot apply the phase rotation trick to the problem since a single phase rotation cannot simultaneously make $\bg_i^H \bff^{(\rm{s})}$ real and nonnegative for all reference locations.
To recast the non-convex SINR expressions into convex forms, we redefine the optimization variables as matrices, i.e., $\bF^{(\rm{s})}=\bff^{(\rm{s})} (\bff^{(\rm{s})})^H, \bF_u^{(\rm{c})}=\bff_u^{(\rm{c})}(\bff_u^{(\rm{c})})^H$, and rewrite the channel vectors as matrices, i.e., $\bG_i = \bg_i \bg_i^H, \bH_u = \bh_u \bh_u^H$. 
The new matrix variables satisfy $\bF^{(\rm{s})}\succeq\bm{0}$ and $\bF_u^{(\rm{c})}\succeq\bm{0}$, together with $\operatorname{rank}(\bF^{(\rm{s})})=1$ and $\operatorname{rank}(\bF_u^{(\rm{c})})=1$. We apply the semidefinite relaxation (SDR)~\cite{Luo2010} by dropping the rank-one constraints while retaining the positive-semidefinite constraints.
With these definitions, we can rewrite the SINR at the tag and the access point as follows:
\begin{equation}
	\rm{SINR}_{i}^{(\rm{t})}(\bF^{(\rm{s})}, \{\bF_u^{(\rm{c})}\}_{u=1}^{U})= \frac{\rm{Tr}(\bG_i \bF^{(\rm{s})})}{\sum_{u=1}^{U}\rm{Tr}(\bG_i \bF_u^{(\rm{c})}) + \sigma_\rm{t}^2},
\end{equation}
\begin{gather}
	\begin{aligned}
		&\rm{SINR}_{i}^{(\rm{r})}(\bF^{(\rm{s})}, \{\bF_u^{(\rm{c})}\}_{u=1}^{U}) \\
		&= \frac{\eta |\bw_i^H \bg_i|^2 \rm{Tr}(\bG_i \bF^{(\rm{s})})}{\sum_{u=1}^{U} \eta |\bw_i^H \bg_i|^2 \rm{Tr}(\bG_i \bF_u^{(\rm{c})}) + \eta \sigma_\rm{t}^2 |\bw_i^H \bg_i|^2 + \sigma_\rm{r}^2}.
	\end{aligned}
\end{gather}
Similarly, the SINR at the user can be rewritten as presented in \eqref{eq:SINR_user_reformulated}.
\begin{figure*}[!t]
	\begin{equation} \label{eq:SINR_user_reformulated}
		\rm{SINR}_{i,u}^{(\rm{c})}(\bF^{(\rm{s})}, \{\bF_u^{(\rm{c})}\}_{u=1}^{U}) = \frac{\rm{Tr}(\bH_u \bF_u^{(\rm{c})})}{\sum_{l \neq u}^{U} \rm{Tr}(\bH_u \bF_l^{(\rm{c})}) + \rm{Tr}(\bH_u \bF^{(\rm{s})}) + \eta |{h}_{i,u}|^2 \left(\rm{Tr}(\bG_i \bF^{(\rm{s})}) + \sum_{v=1}^{U} \rm{Tr}(\bG_i \bF_v^{(\rm{c})}) + \sigma_\rm{t}^2\right)+\sigma_\rm{u}^2}.
	\end{equation}
	\hrule
\end{figure*}
 Then, for compactness, we define the following affine interference-plus-noise terms for the tag-side, reader-side, and user SINRs, respectively:
\begin{align}
    \zeta_i^{(\rm{t})}
    &=
    \sum_{u=1}^{U}
    \operatorname{Tr}(\bG_i\bF_u^{(\rm{c})})
    +\sigma_\rm{t}^2,
    \\
    \zeta_i^{(\rm{r})}
    &=
    \kappa_i
    \sum_{u=1}^{U}
    \operatorname{Tr}(\bG_i\bF_u^{(\rm{c})})
    +\kappa_i\sigma_\rm{t}^2+\sigma_\rm{r}^2,
    \\
    \zeta_{i,u}^{(\rm{c})}
    &=
    \sum_{l\neq u}^{U}
    \operatorname{Tr}(\bH_u\bF_l^{(\rm{c})})
    +\operatorname{Tr}(\bH_u\bF^{(\rm{s})})
    \\
    &
    +\eta|h_{i,u}|^2 \!
    \left(
        \operatorname{Tr}(\bG_i\bF^{(\rm{s})})
        +
        \sum_{v=1}^{U}
        \operatorname{Tr}(\bG_i\bF_v^{(\rm{c})})
        +\sigma_\rm{t}^2
    \right) +\sigma_\rm{u}^2, \nonumber
\end{align}
where $\kappa_i=\eta|\bw_i^H\bg_i|^2$.
Using the strictly positive SINR denominators, the fractional SINR constraints can be equivalently written in cross-multiplied form. After dropping the rank-one constraints, the resulting semidefinite-relaxed mixed-integer problem is formulated as
\begin{IEEEeqnarray}{cl}
\label{eq:opt_BF_SDR}
    \min_{\bF^{(\rm{s})},\{\bF_u^{(\rm{c})}\}_{u=1}^{U},\by}
    &-\displaystyle\sum_{i=1}^{|\cG|}[\by]_i
    \nonumber\\
    \textrm{s.t.}
    &\gamma_\rm{t}[\by]_i \zeta_i^{(\rm{t})}
    -\operatorname{Tr}(\bG_i\bF^{(\rm{s})})
    \leq 0,\quad \forall i \in \mathcal{I},
    \nonumber\\
    &\gamma_\rm{r}[\by]_i \zeta_i^{(\rm{r})}
    -\kappa_i\operatorname{Tr}(\bG_i\bF^{(\rm{s})})
    \leq 0,\quad \forall i \in \mathcal{I},
    \nonumber\\
    &\gamma_\rm{u}\zeta_{i,u}^{(\rm{c})}
    -\operatorname{Tr}(\bH_u\bF_u^{(\rm{c})})
    \leq 0,\quad \forall i \in \mathcal{I}, u \in \mathcal{U},
    \nonumber\\
    &\operatorname{Tr}(\bF^{(\rm{s})})
    +\displaystyle\sum_{u=1}^{U}
    \operatorname{Tr}(\bF_u^{(\rm{c})})
    \leq P,
    \nonumber\\
    &\bF^{(\rm{s})}\succeq\bm{0},\quad
    \bF_u^{(\rm{c})}\succeq\bm{0},\quad \forall u \in \mathcal{U},
    \nonumber\\
    &[\by]_i\in\{0,1\},\quad \forall i \in \mathcal{I}.
\end{IEEEeqnarray}
The first two constraint families contain products between the binary variables $[\by]_i$ and affine functions of the covariance matrices. For any fixed binary vector $\by$, all constraints are affine in the covariance matrices, together with positive-semidefinite cone constraints. Hence, the continuous subproblem is a convex semidefinite program. Moreover, for fixed covariance matrices, these constraints are affine in the binary variables. This structure enables the application of generalized Benders decomposition.
We rewrite the problem as a minimization problem, with the objective function being the negative of the original objective function to align with the default convention of generalized Benders decomposition.

\subsection{Generalized Benders Decomposition}
Generalized Benders decomposition \cite{Geoffrion1972} decomposes an MINLP problem into two subproblems: (i) a \textit{primal problem} with continuous optimization variables and (ii) a \textit{master problem} with discrete optimization variables. The two subproblems are iteratively solved until convergence.

\textbf{Primal Problem.} In each iteration, the primal problem is defined by fixing the discrete variables, i.e., $\by^{(v)}$, where $v$ denotes the iteration counter. For ease of notation, the constraints of the problem in \eqref{eq:opt_BF_SDR} can be integrated as
\begin{gather} 
	\begin{aligned}[b]
	&\bc(\bF^{(\rm{s})}, \{\bF_u^{(\rm{c})}\}_{u=1}^{U}, \by^{(v)})  \\
	&=
		\begin{bmatrix*}[l]
			\gamma_\rm{t} \, [\by]_i^{(v)} \,
			\zeta_i^{(\rm{t})}
			-\operatorname{Tr}(\bG_i\bF^{(\rm{s})}),
			\quad \forall i \in \mathcal{I}
			\\
			\gamma_\rm{r} \, [\by]_i^{(v)} \,
			\zeta_i^{(\rm{r})}
			-\kappa_i\operatorname{Tr}(\bG_i\bF^{(\rm{s})}),
			\quad \forall i \in \mathcal{I}
			\\
			\gamma_\rm{u} \, \zeta_{i,u}^{(\rm{c})}
			-\operatorname{Tr}(\bH_u\bF_u^{(\rm{c})}),
			\quad \forall i \in \mathcal{I}, u \in \mathcal{U}
			\\
			\operatorname{Tr}(\bF^{(\rm{s})})
			+\displaystyle\sum_{u=1}^{U}
			\operatorname{Tr}(\bF_u^{(\rm{c})})-P
		\end{bmatrix*}_{D\times1},
	\end{aligned}
\end{gather}
where $D = 2|\cG| + U|\cG| + 1 $ denotes the dimension. Since the optimization variables $\bF^{(\rm{s})}, \{\bF_u^{(\rm{c})}\}_{u=1}^{U}$ are not in the objective function of \eqref{eq:opt_BF_SDR}, we cannot formulate a standard primal problem. Instead, we introduce slack variables $\bm{\alpha} \in \bbR^{D\times1}$, and the modified primal problem can be formulated as
\begin{gather} 
	\begin{aligned} \label{eq:primal_problem}
		\min_{\bF^{(\rm{s})}, \{\bF_u^{(\rm{c})}\}_{u=1}^{U}, \bm{\alpha}} \quad & \sum_{d=1}^{D}[\bm{\alpha}]_d \\
		\textrm{s.t.} \quad & \bc(\bF^{(\rm{s})}, \{\bF_u^{(\rm{c})}\}_{u=1}^{U}, \by^{(v)}) \leq \bm{\alpha}, \\
		\quad & [\bm{\alpha}]_d \geq 0, \ \forall d \in \{1,\ldots,D\}, \\
		\quad & \bF^{(\rm{s})}\succeq\bm{0},\quad \bF_u^{(\rm{c})}\succeq\bm{0}, \quad \forall u \in \mathcal{U}.
	\end{aligned}
\end{gather}
In each iteration, we obtain the solution $\bF^{(\rm{s})(v)}, \{\bF_u^{(\rm{c})(v)}\}_{u=1}^{U}$ and the corresponding Lagrange multipliers $\bm{\lambda}^{(v)} \in \bbR^{D\times1}$. If the original primal problem is feasible, then $\sum_{d=1}^{D}[\bm{\alpha}]_d = 0$; otherwise, the original primal problem is not feasible. Consequently, we define $\cV_{\rm F}$ and $\cV_{\rm I}$ as the sets of iteration counters associated with feasible and infeasible primal problems, respectively. Then, the Lagrange function is given by
\begin{align}
	&\xi(\bF^{(\rm{s})(v)}, \{\bF_u^{(\rm{c})(v)}\}_{u=1}^{U}, \by, \bm{\lambda}^{(v)}) \\
	& = 
	\begin{dcases}
		-\sum_{i=1}^{|\cG|} [\by]_i + \bm{\lambda}^{(v)T} \bc(\bF^{(\rm{s})(v)}, \{\bF_u^{(\rm{c})(v)}\}_{u=1}^{U}, \by), & v \in \cV_{\rm F}, \\
		\bm{\lambda}^{(v)T} \bc(\bF^{(\rm{s})(v)}, \{\bF_u^{(\rm{c})(v)}\}_{u=1}^{U}, \by), & v \in \cV_{\rm I}. \nonumber
	\end{dcases}
\end{align}

\textbf{Relaxed Master Problem.} 
The master problem is derived from \eqref{eq:opt_BF_SDR} using non-linear convex duality theory \cite{Geoffrion1972} but contains infinitely many constraints. It is therefore relaxed by retaining only a few of the constraints, while the primal problem checks the feasibility of the resulting solution and iteratively adds cutting planes as constraints to reduce the feasible region. The relaxed master problem is formulated as
\begin{subequations} \label{eq:master_problem}
	\begin{align} 
		\min_{\by, \omega} \quad & \omega \\
		\textrm{s.t.} \quad & \omega \geq \xi(\bF^{(\rm{s})(v)}, \{\bF_u^{(\rm{c})(v)}\}_{u=1}^{U}, \by, \bm{\lambda}^{(v)}), \!\! \ \forall v \in \cV_{\rm F}, \label{eq:MP_constraint_1}\\
		\quad & 0 \geq \xi(\bF^{(\rm{s})(v)}, \{\bF_u^{(\rm{c})(v)}\}_{u=1}^{U}, \by, \bm{\lambda}^{(v)}), \ \forall v \in \cV_{\rm I}, \label{eq:MP_constraint_2}\\
		\quad & [\by]_i \in \{0, 1\}, \ \forall i \in \mathcal{I},
	\end{align}
\end{subequations}
where $\omega$ is an auxiliary optimization variable. \eqref{eq:MP_constraint_1} and \eqref{eq:MP_constraint_2} are the optimality and feasibility cuts, respectively.

\textbf{Initialization and Convergence.}
To initialize the algorithm, a feasible solution to \eqref{eq:opt_BF_SDR} is required to generate the initial optimality cut, preventing the relaxed master problem \eqref{eq:master_problem} from being unbounded. We select $[\by]_i^{(v)} =0, \forall i \in \mathcal{I}$ as the initial solution, which renders the tag-side and reader-side SINR constraints satisfied. This initialization is feasible provided that the communication user SINR requirements can be satisfied under the total power budget.
In each iteration, the relaxed master problem, which involves only a subset of the constraints, provides a lower bound (LBD) on the objective value of \eqref{eq:opt_BF_SDR}. The candidate solution $\by^{(v)}$ from the relaxed master problem is tested in the modified primal problem. If feasible, the primal problem yields the upper bound (UBD), updated monotonically to track the best-known value. The process terminates when the gap between UBD and LBD satisfies $\rm{UBD}-\rm{LBD} \leq \epsilon$ for tolerance $\epsilon$.

\textbf{Complexity Analysis.} 
Assuming convergence is achieved in $K$ iterations, the algorithm requires solving $K$ semidefinite programs (the primal problem) and $K$ integer linear programs (the relaxed master problem). Thus, the computational complexity is dominated by the NP-hard relaxed master problem, which may grow exponentially with the grid size $|\cG|$ in the worst case. However, in the context of our ISAC application, this complexity is manageable due to the static nature of the scenarios. For instance, in a warehouse, the access point and communication users (e.g., surveillance cameras) are fixed. This allows for offline pre-computation of the codebook during the planning stage based on tag interrogation coverage.

\begin{algorithm}[t]
	\caption{Site-Specific Beamforming Codebook Design}
	\label{alg1}
	\begin{algorithmic}[1]
		
		\Require \begin{varwidth}[t]{\columnwidth} 
			$\cF=\emptyset$, $\Theta_{\rm{min}}=0$, $\Theta_{\rm{max}}=\Theta_{\rm{step}}$.
		\end{varwidth}
		
		\While{$\Theta_{\rm{min}} < 180^\circ$}  

		\State Select the channel records $\mathcal{D}_{\mathcal{S}}$ for the current sector.

		\State Initialize the generalized Benders decomposition: $\rm{UBD}=0$, $\rm{LBD}=-|\cG|$, $\cV_{\rm F}=\emptyset$, $\cV_{\rm I}=\emptyset$, $v=0$.

		\State Select an initial feasible solution to \eqref{eq:opt_BF_SDR}: $\by^{(v)}=\bm{0}$.

		\State Solve the modified primal problem \eqref{eq:primal_problem}, and obtain $\bF^{(\rm{s})(v)}, \{\bF_u^{(\rm{c})(v)}\}_{u=1}^{U}$, and $\bm{\lambda}^{(v)}$.

		\State Update $\cV_{\rm F} = \cV_{\rm F} \cup \{v\}$, $\bF_{\rm{best}}^{(\rm{s})}=\bF^{(\rm{s})(v)}$ and $\bF_{\rm{best},u}^{(\rm{c})} =\bF_u^{(\rm{c})(v)}$, $\forall u \in \mathcal{U}$.

		\While{$\rm{UBD}-\rm{LBD} > \epsilon$}

		\State Set $v = v + 1$.

		\State Solve the relaxed master problem \eqref{eq:master_problem}, and obtain the optimal solution $\by^{(v)}$ and $\omega^{(v)}$.

		\State Update $\rm{LBD}=\omega^{(v)}$.

		\State Solve the modified primal problem \eqref{eq:primal_problem}, and obtain $\bF^{(\rm{s})(v)}, \{\bF_u^{(\rm{c})(v)}\}_{u=1}^{U}$, and $\bm{\lambda}^{(v)}$.

		\If{$\sum_{d=1}^{D}[\bm{\alpha}]_d = 0$}

		\State Update $\cV_{\rm F} = \cV_{\rm F} \cup \{v\}$.

		\If{$-\sum_{i=1}^{|\mathcal{G}|}[\by]_i^{(v)}<\rm{UBD}$}
		
		\State Update $\rm{UBD}= -\sum_{i=1}^{|\cG|} [\by]_i^{(v)}$.

		\State Store
					$\bF_{\rm{best}}^{(\rm{s})}=\bF^{(\rm{s})(v)}$ and
					$\bF_{\rm{best},u}^{(\rm{c})} =\bF_u^{(\rm{c})(v)}$, $\forall u \in \mathcal{U}$.

		\EndIf

		\Else

		\State Update $\cV_{\rm I} = \cV_{\rm I} \cup \{v\}$.

		\EndIf

		\EndWhile

		\State Apply rank-one approximation to $\bF_{\rm{best}}^{(\rm{s})}, \{\bF_{\rm{best},u}^{(\rm{c})}\}_{u=1}^{U}$, and obtain $\hat{\bff}^{(\rm{s})}, \{\hat{\bff}_u^{(\rm{c})}\}_{u=1}^{U}$.

		\State Update $\cF = \cF \cup \{(\hat{\bff}^{(\rm{s})}, \{\hat{\bff}_u^{(\rm{c})}\}_{u=1}^{U})\}$.

		\State Update $\Theta_{\rm{min}}=\Theta_{\rm{max}}$, $\Theta_{\rm{max}}=\min(\Theta_{\rm{max}}+\Theta_{\rm{step}}, 180^\circ)$.
		
		\EndWhile

		\Ensure Beamforming codebook $\cF$. 
		
	\end{algorithmic}
\end{algorithm}

\subsection{Overall Algorithm for Codebook Construction}
The complete site-specific codebook design is summarized in Algorithm~\ref{alg1}. For each sector, the corresponding channel records are extracted from the offline channel database. The generalized Benders decomposition method then designs a codeword that maximizes the number of successfully interrogated reference locations in that sector. Repeating this procedure for all sectors produces the complete beamforming codebook.
Assuming a sector width of $\Theta_{\rm{step}}$ degrees over a semicircular area, the resulting codebook has a cardinality of $\ceil[\big]{\frac{180}{\Theta_{\rm{step}}}}$. Since the semidefinite relaxation technique is employed, the algorithm yields covariance matrices $\hat{\bF}^{(\rm{s})}, \{\hat{\bF}_u^{(\rm{c})}\}_{u=1}^{U}$ rather than beamforming vectors. If these matrices are rank-one, the beamforming vectors $\hat{\bff}^{(\rm{s})}, \{\hat{\bff}_u^{(\rm{c})}\}_{u=1}^{U}$ can be directly extracted via decomposition. In the general case where the matrices are not rank-one, we extract the candidate vectors using a rank-one approximation based on the dominant eigenvector~\cite{Luo2010}. Specifically, we perform an eigendecomposition on $\hat{\bF}^{(\rm{s})}$ as follows:
\begin{equation}
	\hat{\bF}^{(\rm{s})} = \sum_{m=1}^{M} \lambda_m \bu_m \bu_m^H,
\end{equation}
where $\lambda_1 \geq \lambda_2 \geq \cdots \geq \lambda_M \geq 0$ are the eigenvalues of $\hat{\bF}^{(\rm{s})}$, and $\bu_m$ denotes the eigenvector corresponding to $\lambda_m$. We approximate the beam using the principal component $\bu_1$, scaled by the allocated transmit power, i.e., $\hat{\bff}^{(\rm{s})} = \sqrt{\rm{Tr}(\hat{\bF}^{(\rm{s})})} \bu_1$, where $\rm{Tr}(\hat{\bF}^{(\rm{s})})$ represents the allocated transmit power. The communication beams $\{\hat{\bff}_u^{(\rm{c})}\}_{u=1}^{U}$ can be obtained similarly.

\section{Simulation Results} \label{sec:simulation_results}

\subsection{Simulation Setup}
We consider a simulation environment where the access point is located at the origin of the Cartesian coordinate system. The shared antenna array of the access point is a uniform linear array aligned along the y-axis, with the boresight directed toward the positive x-axis. The system operates at a carrier frequency of $915$ MHz with a half-wavelength antenna spacing. The total transmit power of the access point is set to $P=30$ dBm. Unless otherwise stated, we consider a single communication user and line-of-sight (LoS) propagation for all wireless links. The communication user SINR requirement is set to 10 dB. The receiver noise power is calculated as $\sigma_\rm{t}^2=\sigma_\rm{r}^2=\sigma_\rm{u}^2=10\log_{10}(kT_0B)+N_f+30$ dBm, where $k$ is Boltzmann's constant, $T_0=290$ kelvin, and $B=10$ MHz. We assume a noise figure of $N_f=7$ dB for the reader and the user and $N_f=0$ dB for the passive tag. Based on the commercial datasheets~\cite{Impinj_Tag,Impinj_Reader}, the sensitivity values are set to $P_{\rm{sen,t}} = -25.5$ dBm for the tag and $P_{\rm{sen,r}} = -94$ dBm for the reader.
The backscatter-modulation efficiency of the tag is set to $\eta = 0.16$, assuming a given differential radar cross section~\cite{Impinj_Tag} and an FM0 encoding scheme~\cite{Nikitin2008}. Finally, the modified primal problem in \eqref{eq:primal_problem} and the relaxed master problem in \eqref{eq:master_problem} are solved using the MOSEK solver \cite{mosek} within the CVX framework \cite{cvx}.

\subsection{Performance Evaluation of the Beamforming Design}
\textbf{Which RFID SINR constraint dominates the interrogation process?}
Before evaluating the beamforming designs, we investigate whether the tag-side or reader-side SINR constraint acts as the limiting factor in the interrogation process, as discussed in Proposition 1. We analyze this by varying the tag location under two antenna array sizes (four and eight antennas). Table \ref{tab:dominant_link} summarizes the dominant constraint and its associated interrogation distance range. The tag-side constraint dominates from zero meters up to a certain distance, beyond which the reader-side constraint takes precedence. As the number of antennas increases, the range of the tag-side-dominated region expands. Notably, as will be discussed in the subsequent evaluation, the achievable interrogation distance primarily falls within the tag-side-dominated region. This observation underscores the critical role of transmit beamforming design in enhancing the received signal strength at the tag.

\begin{figure}[!t]
	\begingroup
	\centering
	
	\includegraphics[width=1.0\columnwidth]{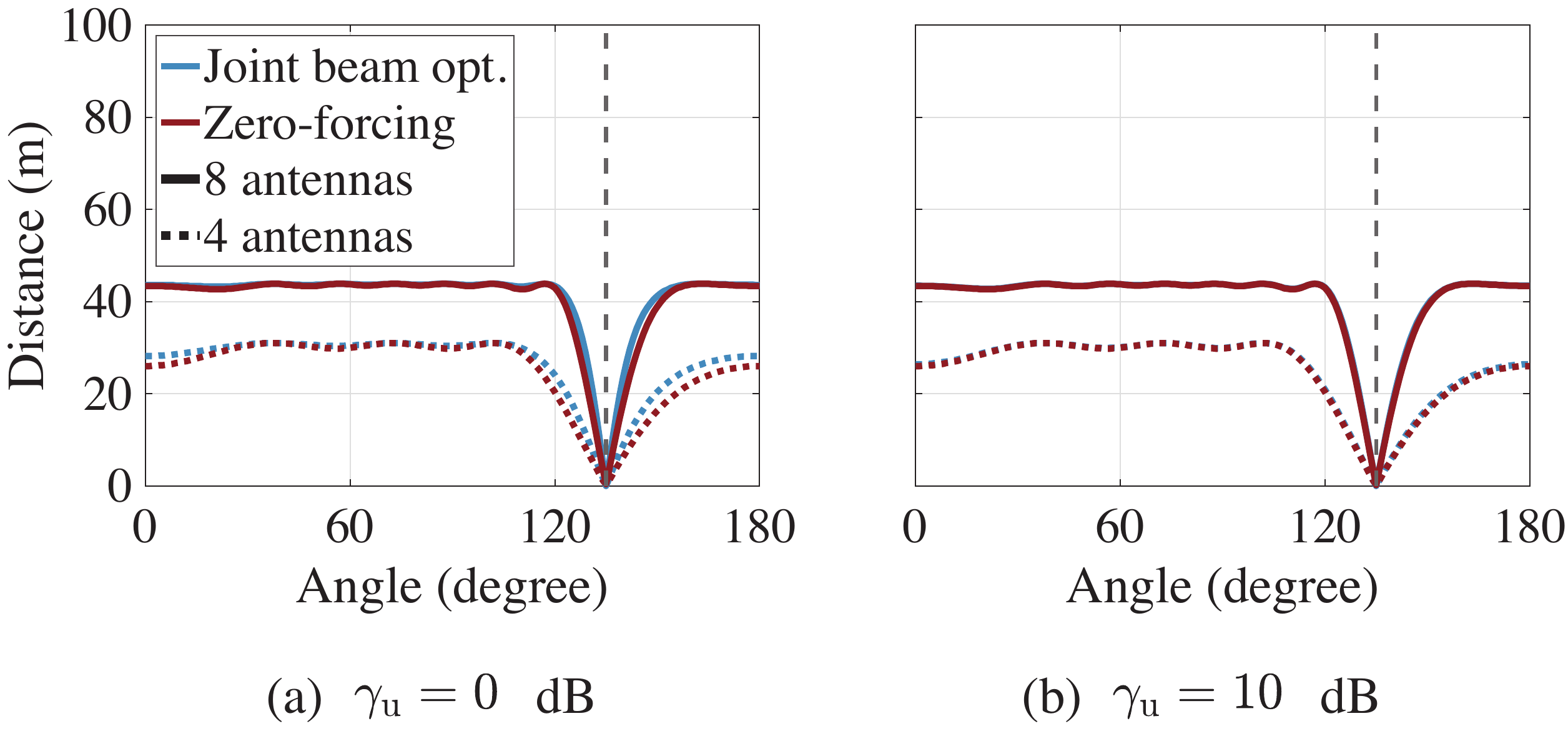}
	
	\caption{Achievable interrogation distance versus tag direction for a communication user located $5$ meters from the access point at $135^{\circ}$. Subplots (a) and (b) compare the proposed beamforming designs under different user SINR requirements. More antennas increase the interrogation distance, with a dip when the tag and user are directionally aligned. Under a low SINR requirement, joint optimization outperforms zero forcing by flexibly balancing tag enhancement and user interference.}
	\label{fig:max_distance}
	\par\endgroup
	\vspace{\floatsep}
	\begingroup
	\centering	

	\includegraphics[width=0.98\columnwidth]{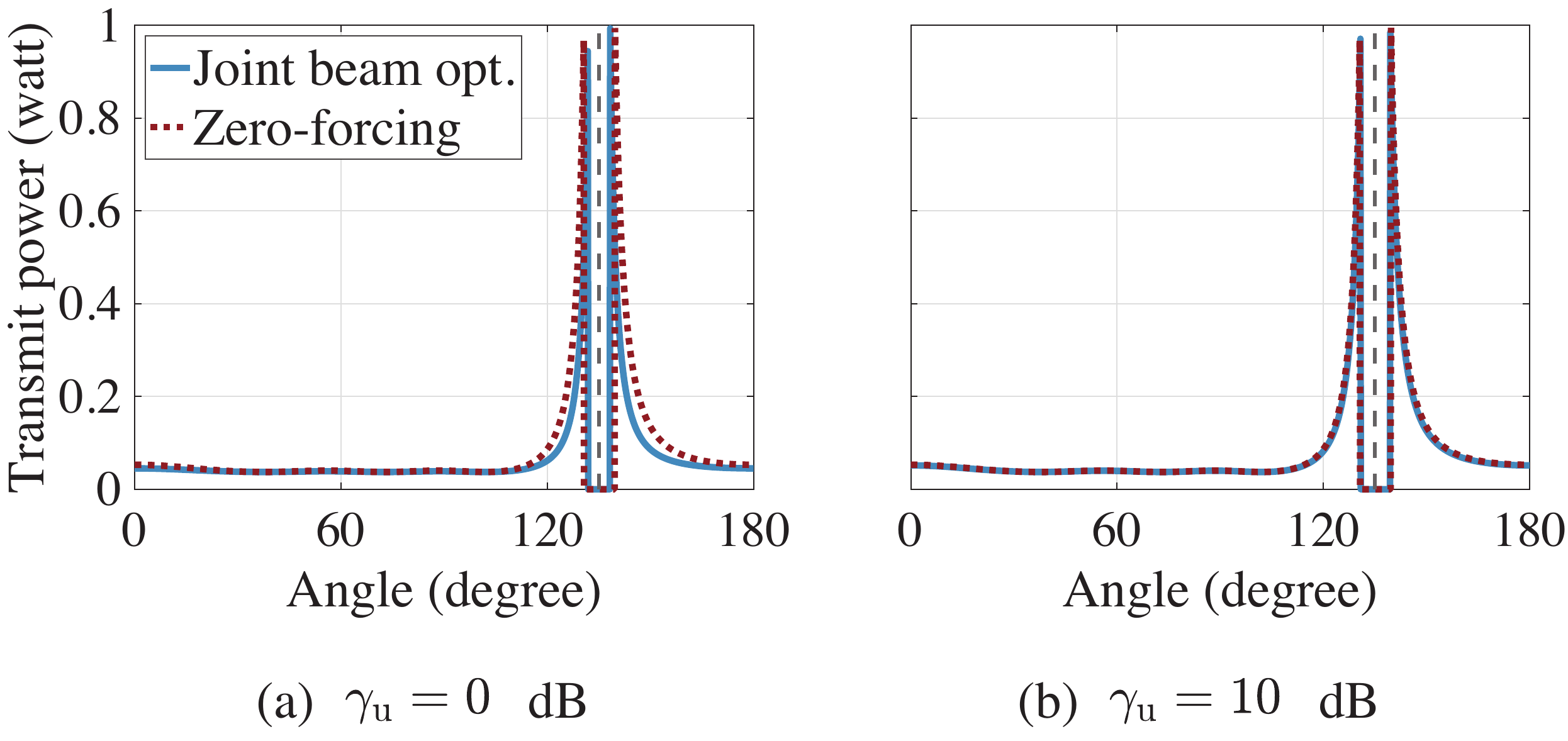}

	\caption{Total allocated transmit power versus tag direction with four antennas. The tag is $6$ meters from the access point, and the communication user is at $135^{\circ}$ and $5$ meters. Subplots (a) and (b) compare the joint beamforming and zero-forcing methods for user SINR requirements of $0$ dB and $10$ dB, respectively. Power rises sharply as the tag and user directions align due to stronger interference. The joint design consumes less power, particularly at low SINR, owing to its greater flexibility in managing interference.}	
	\label{fig:transmit_power}
	\par\endgroup
\end{figure}

\textbf{What is the achievable interrogation distance, and how does the number of antennas affect it?}
We evaluate the maximum achievable interrogation distance as a function of the tag angular direction. In this simulation, we place the tag at various angles ranging from $0^{\circ}$ to $180^{\circ}$. For each angle, we incrementally increase the tag distance until the optimization problem becomes infeasible. The communication user is fixed at $135^\circ$ and 5 meters from the access point. \figref{fig:max_distance} shows the results for two antenna configurations (four and eight antennas) under different user SINR requirements (0 dB and 10 dB). Increasing the number of antennas significantly enhances the interrogation range due to higher beamforming gain. Specifically, the maximum distances for four and eight antennas are up to 31.0 and 43.9 meters, respectively. When the tag and user are directionally aligned, a performance dip is observed due to the increased interference. Comparing the methods, \figref{fig:max_distance}(a) shows that the joint beamforming optimization outperforms the zero-forcing method under a low user SINR requirement, where the user can tolerate more interference. This is because the joint design can flexibly balance between enhancing the tag signal and controlling interference, whereas the zero-forcing method rigidly eliminates interference. Conversely, under a high user SINR requirement, both methods exhibit comparable performance, as shown in \figref{fig:max_distance}(b), which indicates that effective interference mitigation becomes crucial for both approaches.

\begin{table}[!t]
	\renewcommand{\arraystretch}{1.3}
	\caption{Constraint-dominance regions.}
	\label{tab:dominant_link}
	\centering
	\begin{tabular}{c c c}
	\toprule
	Number of antennas & \makecell{Tag-side-dominated \\ region} & \makecell{Reader-side-dominated \\ region} \\
	\midrule
	4 & $0 - 55.50$ meters & $> 55.50$ meters \\
	\midrule
	8 & $0 - 78.49$ meters & $> 78.49$ meters \\
	\bottomrule
	\end{tabular}
\end{table}

\textbf{What is the total allocated power of the proposed solutions, and what is the impact of mutual interference?}
We analyze the total allocated transmit power, which is the primary objective of our optimization problems, using four antennas. The tag is placed at a fixed radius of 6 meters with an angular direction varying from $0^{\circ}$ to $180^{\circ}$. The communication user is at $135^{\circ}$ and $5$ meters. \figref{fig:transmit_power} presents the power consumption for both methods under 0 dB and 10 dB user SINR constraints; a zero power value in the figure indicates that no feasible solution exists. As the tag angle approaches the user angle, the required transmit power escalates until reaching the maximum budget, driven by the need to overcome increased mutual interference. A key distinction appears under low user SINR requirements, where the joint beamforming optimization consistently consumes less power than the zero-forcing method, as shown in \figref{fig:transmit_power}(a). This efficiency stems from the flexibility of the joint design to manage interference rather than strictly nulling it. In \figref{fig:transmit_power}(b), both methods converge to similar power consumption patterns under high user SINR requirements, which indicates that interference suppression becomes the dominant factor for both approaches.

\begin{figure}[!t]
	\begingroup
	\centering	

	\includegraphics[width=0.995\columnwidth]{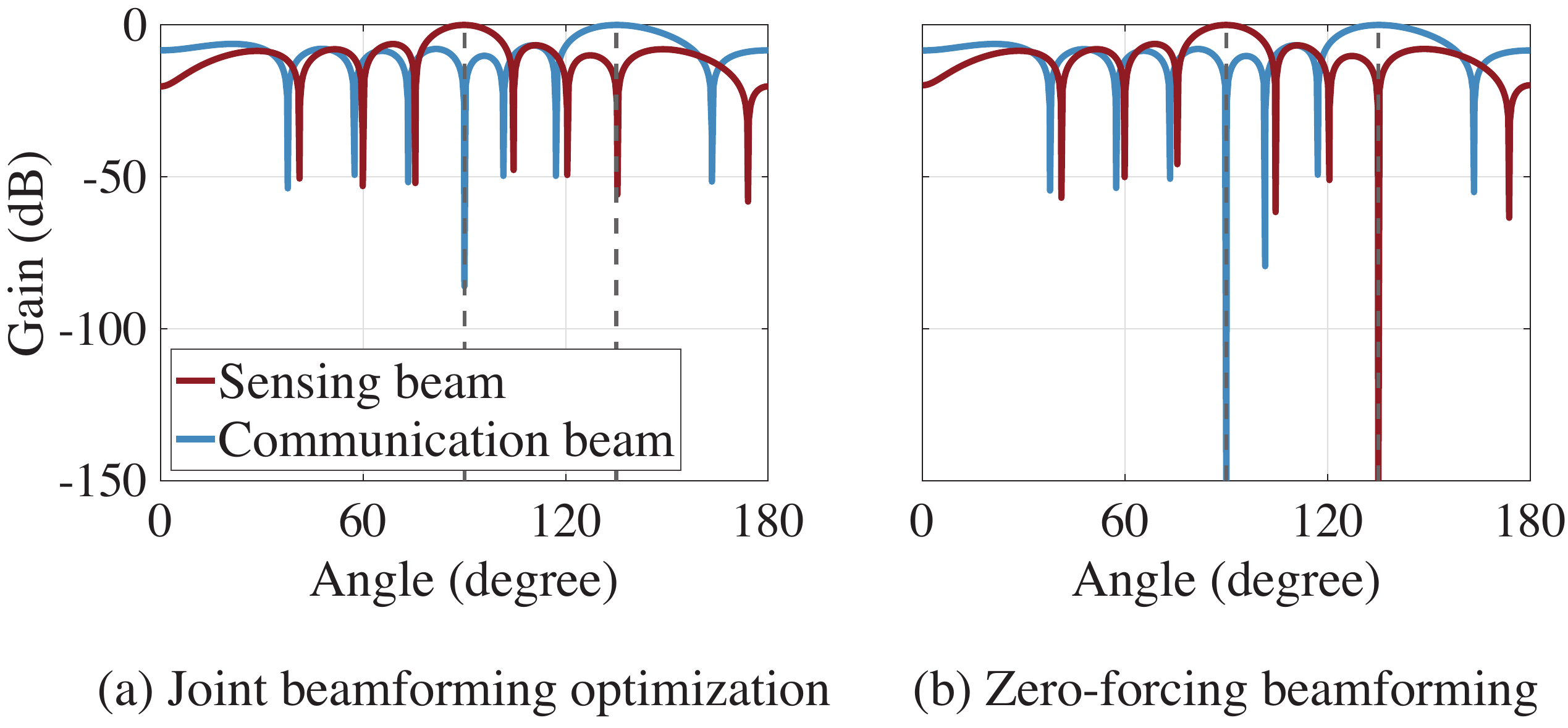}

	\caption{Normalized beamforming patterns for the sensing and communication beams with eight antennas. The tag is located at $90^{\circ}$, while the communication user is positioned at $135^{\circ}$. (a) The joint beamforming design optimizes the null depth based on SINR requirements. By allowing shallower nulls, the joint design preserves degrees of freedom to enhance the main lobe gain toward the tag. (b) The zero-forcing method enforces deep, strict nulls toward the interference directions, which consumes degrees of freedom.}	

	\label{fig:beam_pattern}
	\par\endgroup
	\vspace{\floatsep}
	\begingroup
	\centering

	\includegraphics[width=0.95\columnwidth]{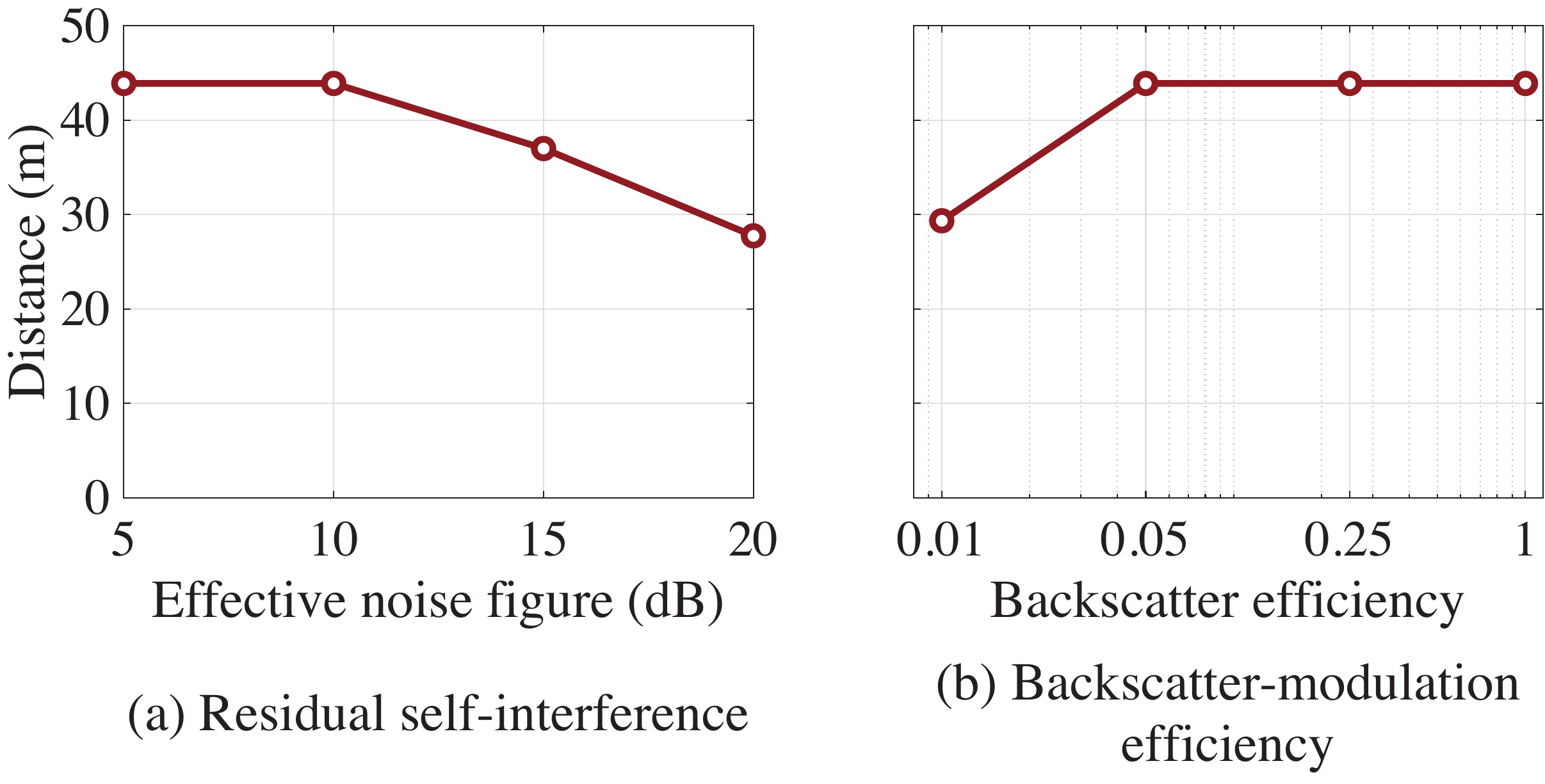}

	\caption{Maximum achievable interrogation distance over tag directions versus (a) the residual self-interference and (b) the tag backscatter-modulation efficiency. The access point employs eight antennas, while the communication user is at $135^{\circ}$ and $5$ meters. The results show that the interrogation distance decreases under sufficiently strong residual interference or low backscatter efficiency and is otherwise limited by the tag-side constraint.}
	\label{fig:max_distance_vs_nf_eta}
	\par\endgroup
\end{figure}

\textbf{How do the beamforming patterns reveal the interference mitigation capability of the proposed solutions?}
To visualize the interference mitigation capabilities of the proposed methods, we examine their respective beamforming patterns. The tag and communication user are positioned at angles of $90^{\circ}$ and $135^{\circ}$, respectively, and the access point is equipped with eight antennas. \figref{fig:beam_pattern} shows the normalized beamforming patterns. In both approaches, the main lobes are steered toward their intended targets, while nulls are formed in the interference directions; specifically, the sensing beam suppresses the communication user, while the communication beam suppresses the tag. The joint beamforming design in \figref{fig:beam_pattern}(a) adapts the null depths according to the SINR requirements and therefore permits shallower nulls. This flexibility preserves more spatial degrees of freedom and enables higher beamforming gain toward the tag, explaining its superior interrogation distance and power efficiency observed in the previous simulations. In comparison, the zero-forcing method in \figref{fig:beam_pattern}(b) enforces deep nulls to eliminate cross-interference, consuming more spatial degrees of freedom that could otherwise be used to enhance the desired signals.

\textbf{How do residual self-interference and backscatter-modulation efficiency affect the interrogation distance?} For this system-level study, we do not explicitly model residual transmitter leakage or cancellation techniques. Because transmitter self-interference can degrade the talk-mode receiver sensitivity of practical RFID readers~\cite{Khannur2008}, we represent the aggregate post-cancellation impairment as an increase in the effective reader noise figure. Using zero-forcing beamforming with eight antennas, we evaluate the maximum achievable interrogation distance over the tag directions. As shown in \figref{fig:max_distance_vs_nf_eta}(a), the distance remains nearly unchanged for effective noise figures up to 10 dB but decreases under stronger residual interference. Thus, moderate residual interference has limited impact while the tag-side constraint dominates, whereas sufficiently strong interference makes reader-side detection the limiting factor. We also examine the effect of tag backscatter-modulation efficiency, which is influenced by the radar cross section and modulation scheme. As shown in \figref{fig:max_distance_vs_nf_eta}(b), the distance remains at 43.9 meters for efficiencies of 0.05 or higher but decreases to 29.4 meters at an efficiency of 0.01. Thus, the tag-side constraint determines the range once the backscattered signal is sufficiently strong, while low backscatter efficiency makes reader-side detection dominant.

\subsection{Performance Evaluation of the Codebook Design} \label{subsec:codebook_evaluation}
To evaluate the proposed codebook, we consider an access point with eight antennas and a communication user located 5 meters away at $135^{\circ}$. We vary the sector interval $\Theta_{\rm{step}}$ to study the impact of codebook resolution. The reference grid uses $K_r=60$ radial samples with $R_{\rm{min}}=0$ and an angular sampling interval $\Delta_\theta=1^\circ$, or $\Delta_\theta=0.5^\circ$ when $\Theta_{\rm{step}}=1^\circ$. 
To reduce computational overhead, we exclude the locations where no codeword can interrogate a tag, i.e., where even the beams optimized individually for a tag cannot satisfy its SINR requirements. For each sector, we set $R_{\rm{max}}=\max_{\theta\in\vartheta}R_{\rm{max},\theta}$ and keep only the locations with $r\leq R_{\rm{max},\theta}$, where $R_{\rm{max},\theta}$ is the largest distance at which the single-tag joint beamforming problem \eqref{eq:single_tag_opt_BF} remains feasible in direction $\theta$ and is shorter near the communication user due to stronger interference.

We consider two benchmark methods in this study:
\begin{itemize}
	\item \textbf{Point-targeting method.} This method constructs one codeword for each target location by solving the single-tag joint beamforming problem in \eqref{eq:single_tag_opt_BF}. The target directions are sampled from $0^\circ$ to $180^\circ$ with a $1^\circ$ interval, with the tag in each direction placed at its maximum feasible interrogation distance. Each codeword is therefore optimized for one location without explicitly considering coverage over the corresponding sector.
	\item \textbf{Max-min gain method.} This method adapts the SINR-constrained minimum weighted beampattern-gain design in \cite{Hua2023} to the considered RFID system. We apply the max–min criterion only to the sensing covariance matrix and include tag-induced backscatter interference at every reference grid point in the communication user SINR. For each sector, we maximize the minimum sensing beamforming gain evaluated along the sampled angular directions within that sector. Unlike the proposed method, this benchmark uses sensing beamforming gain as a surrogate for interrogation coverage rather than explicitly enforcing the tag-side and reader-side SINR requirements.
\end{itemize}
For a fair comparison, the benchmark codebooks are evaluated using the same tag-side, reader-side, and communication user SINR requirements as the proposed method.

\begin{figure}[!t]
	\centering	

	\includegraphics[width=0.945\columnwidth]{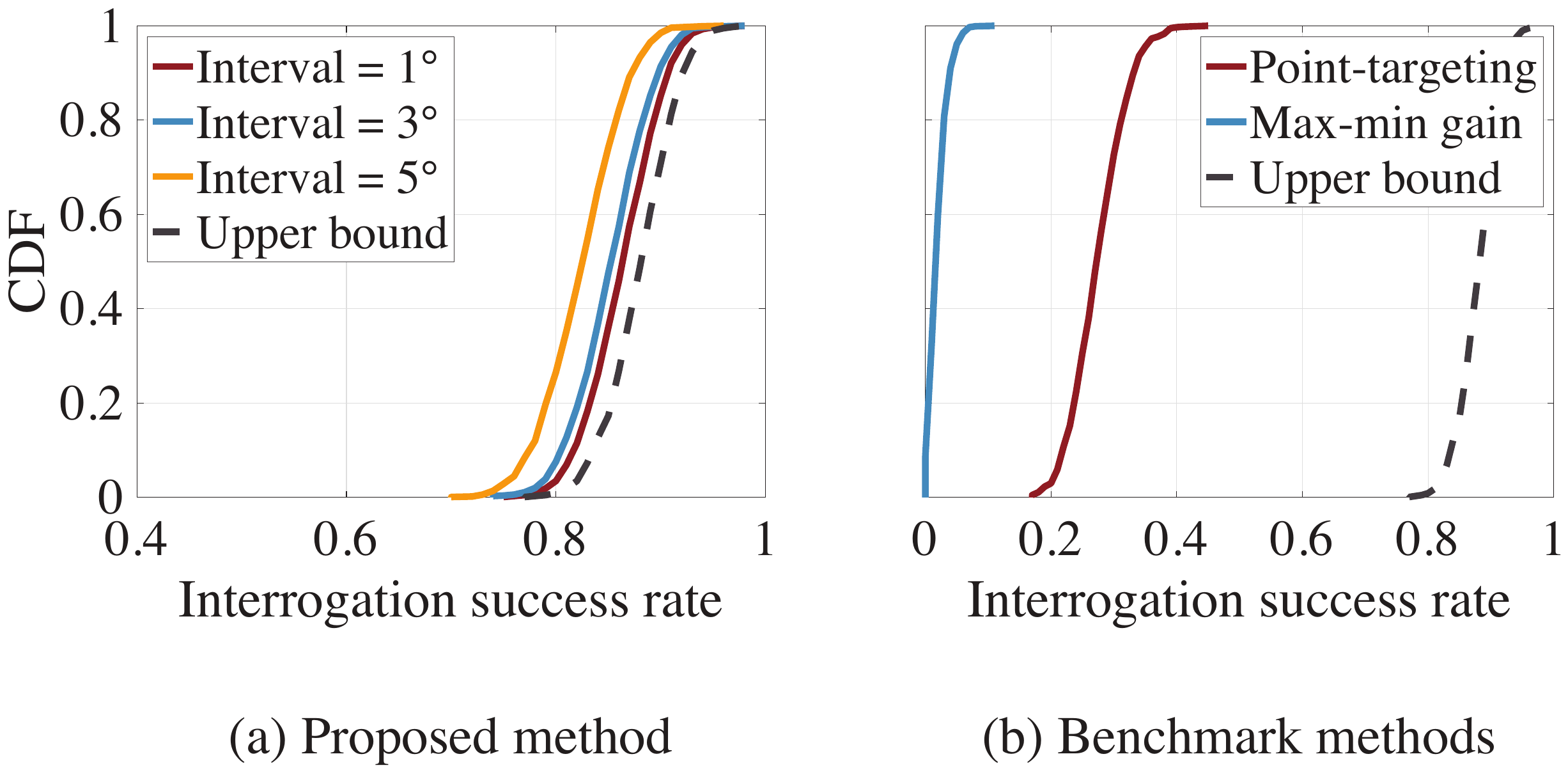}

	\caption{Cumulative distribution function (CDF) of the interrogation success rate. Subplot (a) compares the proposed method with various angle intervals of $1^\circ$, $3^\circ$, and $5^\circ$, while subplot (b) presents the benchmark methods. The results indicate that the proposed method significantly outperforms the benchmarks, achieving higher interrogation success rates.}
	\label{fig:CDF_success_rate}
\end{figure}

\begin{figure}[!t]
	\begingroup
	\centering

	\includegraphics[width=0.97\columnwidth]{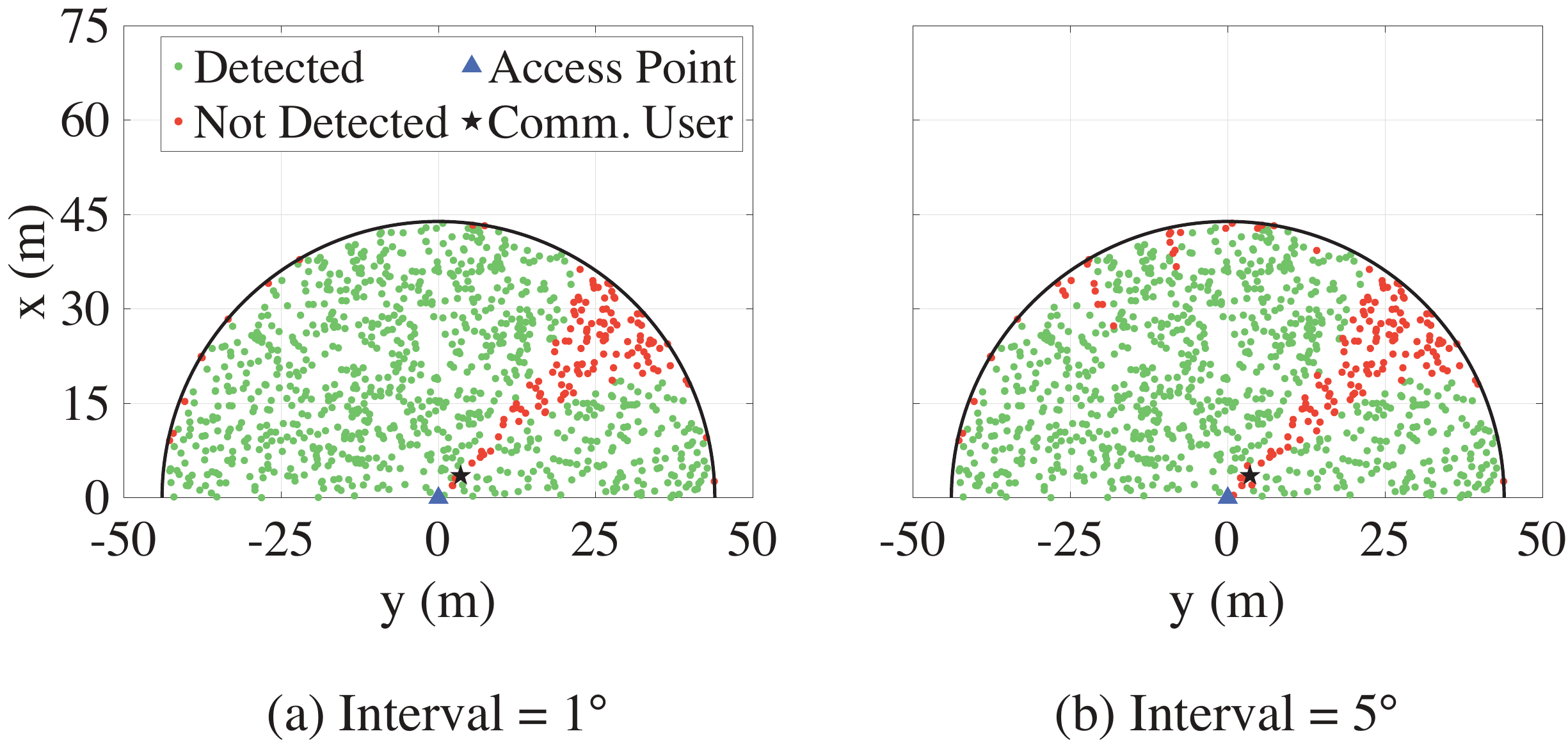}
	\caption{Interrogation coverage maps of the proposed codebook design with sector angle intervals of (a) $1^\circ$ and (b) $5^\circ$. A finer interval provides more comprehensive coverage with fewer uninterrogated tags, whereas a larger interval reduces coverage, particularly near the communication user and at long range, but requires fewer codewords for beam sweeping.}
	\label{fig:coverage}
	\par\endgroup
	\vspace{\floatsep}
	\begingroup
	\centering	

	\includegraphics[width=0.97\columnwidth]{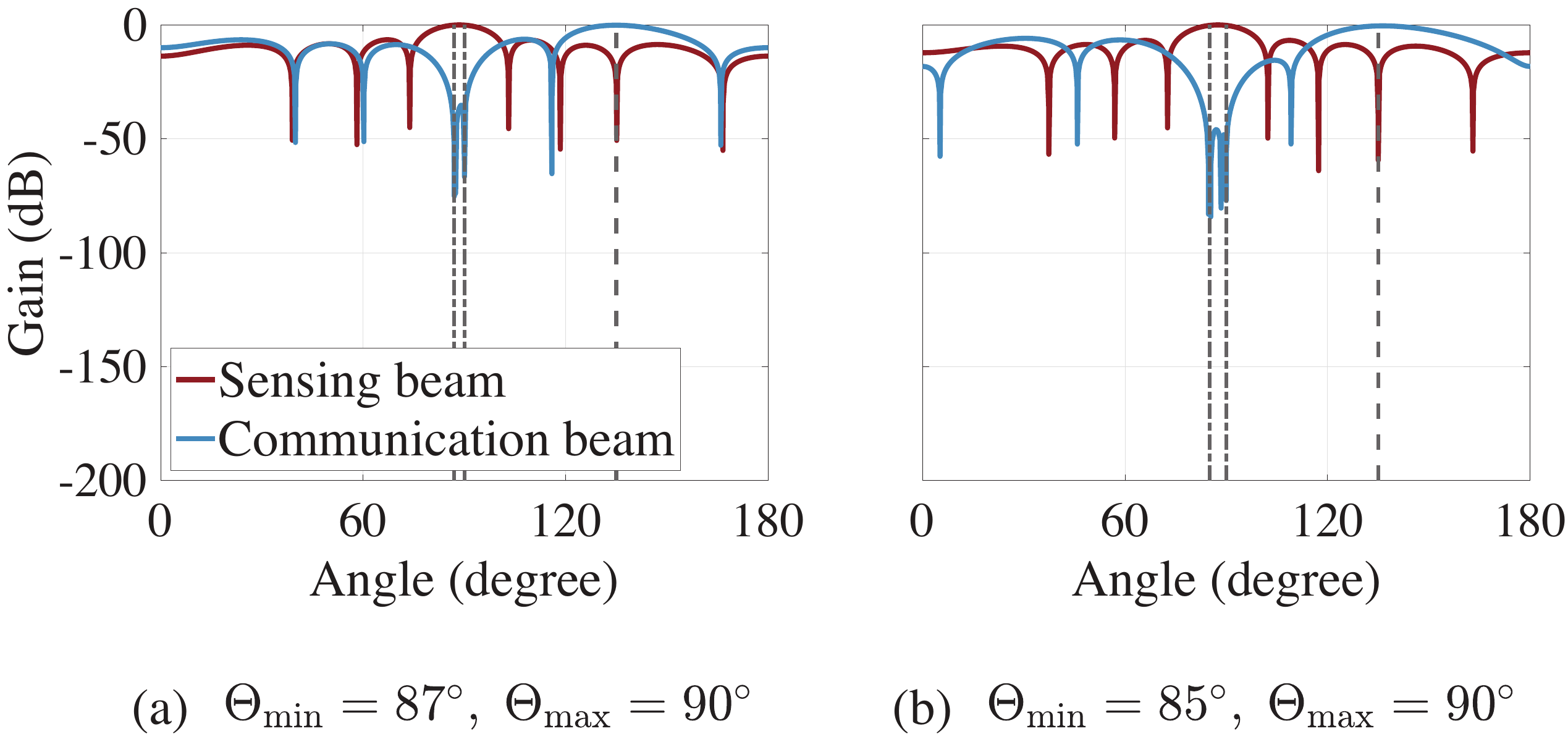}

	\caption{Normalized beamforming patterns with eight antennas and a communication user at $135^\circ$. Subplots (a) and (b) correspond to the $87^\circ$ -- $90^\circ$ and $85^\circ$ -- $90^\circ$ sectors, respectively. The sensing beam nulls the user, while the communication beam suppresses interference throughout each sector. A wider sector requires a broader suppression region.}
	\label{fig:beam_pattern_CB}
	\par\endgroup
\end{figure}

\textbf{What are the interrogation success rate and spatial coverage of the designed codebook?} We define the interrogation success rate as the fraction of tags covered by at least one codeword during a complete beam sweep. We conduct 1000 trials, each involving 100 tags randomly distributed within a semicircular region with a radius of 43.9 meters. \figref{fig:CDF_success_rate}(a) shows that the performance of the proposed method improves as the sector angle interval decreases, with the $1^\circ$ design approaching the upper bound, which is obtained by applying the single-tag joint beamforming design individually to each tag with perfect channel knowledge. \figref{fig:CDF_success_rate}(b) further shows that the proposed method substantially outperforms both benchmarks. The coverage maps in \figref{fig:coverage} show that the $1^\circ$ codebook leaves fewer tags uninterrogated than the $5^\circ$ design, particularly near the communication user and at long range. However, larger sectors require fewer codewords and lower the beam sweeping overhead, revealing a coverage–overhead tradeoff. Since the evaluated tags are randomly distributed and are not restricted to the reference locations, these results also demonstrate that the designed codebook can interrogate tags at arbitrary locations within the region of interest.

\textbf{What do the beamforming patterns and rank-one analysis reveal about the proposed codebook?} To understand the coverage behavior, we examine the normalized beamforming patterns in \figref{fig:beam_pattern_CB}. For both sector intervals, the sensing beam forms a null toward the communication user, while the communication beam suppresses the angular range of the sensing sector. Increasing the sector interval from $3^\circ$ in \figref{fig:beam_pattern_CB}(a) to $5^\circ$ in \figref{fig:beam_pattern_CB}(b) requires a wider suppression region to limit interference with tag interrogation, consuming spatial degrees of freedom and reducing the beamforming gain. This explains the lower coverage for larger sector intervals. We further evaluate the quality and practical effect of the rank-one approximation in Table~\ref{tab:rank_one_quality}. For each interval, the reported rank-one power fractions are averaged over all codewords in the corresponding codebook. The mean rank-one power fractions of the sensing and communication covariance matrices exceed $96\%$ for all considered sector intervals, indicating that the semidefinite-relaxed solutions are close to rank one. Moreover, when evaluated over the reference locations across all sectors, the rank-one approximation retains at least $99.99\%$ of the original constraint satisfaction performance. Therefore, the approximation has a negligible effect on the joint tag-side, reader-side, and communication-user SINR performance.

\begin{table}[!t]
	\begingroup
    \renewcommand{\arraystretch}{1.2}
    \setlength{\tabcolsep}{2.5pt}
    \caption{Rank-one approximation quality.}
    \label{tab:rank_one_quality}
    \centering
    \footnotesize
    \begin{tabular}{c c c c c}
        \toprule
        \makecell{Sector\\interval\\(degrees)}
        &
        \makecell{Codebook\\cardinality\\$|\mathcal{F}|$}
        &
        \makecell{Mean sensing\\rank-one power\\fraction (\%)}
        &
        \makecell{Mean comm.\\rank-one power\\fraction (\%)}
        &
        \makecell{Constraint\\retention\\(\%)}
        \\
        \midrule
        $1$ & $180$ & $98.88$ & $96.02$ &  $99.99$ \\
        $2$ &  $90$ & $98.57$ & $97.38$ & $100.00$ \\
        $3$ &  $60$ & $99.00$ & $98.02$ & $100.00$ \\
        $4$ &  $45$ & $99.08$ & $98.46$ & $100.00$ \\
        $5$ &  $36$ & $98.93$ & $98.49$ &  $99.99$ \\
        \bottomrule
    \end{tabular}
	\par\endgroup
	\vspace{\floatsep}
	\begingroup
    \renewcommand{\arraystretch}{1.2}
    \caption{Mean interrogation success rate (\%) with a $1^\circ$ sector interval under the relaxed model and the adopted condition.}
    \label{tab:relaxed_model}
    \centering
    \footnotesize
    \begin{tabular}{l c c c c c}
        \toprule
        & \multicolumn{4}{c}{Relaxed model, $\gamma_\rm{t}$ (dB)} & Adopted \\
        \cmidrule(lr){2-5}
        & 10 & 20 & 30 & 40 & condition \\
        \midrule
        Codebook    & 88.90 & 87.97 & 87.87 & 87.77 & 86.76 \\
        Upper bound & 89.65 & 88.76 & 88.57 & 88.51 & 88.50 \\
        \bottomrule
    \end{tabular}
	\par\endgroup
\end{table}

\textbf{Does the adopted tag-side condition limit the interrogation success rate?} The tag-side condition in Section~\ref{subsec:backscatter_model} is conservative and may over-suppress the communication beams toward the tags. If the command-decoding SINR threshold is known, the proposed method can be adapted to credit the communication signals toward tag powering by adding the constraint ${\operatorname{Tr}(\bG_i\bF^{(\rm{s})})+\sum_{u=1}^{U}\operatorname{Tr}(\bG_i\bF_u^{(\rm{c})})}\geq P_{\rm{sen,t}}[\by]_i$, $\forall i \in \mathcal{I}$, to \eqref{eq:opt_BF_SDR} and setting $\gamma_\rm{t}$ accordingly. The corresponding upper bound is obtained by adding the counterpart of this constraint for tag $t$ to \eqref{eq:single_tag_opt_BF} and solving the resulting problem via semidefinite relaxation, which yields rank-one solutions in all evaluated cases. As shown in Table~\ref{tab:relaxed_model}, the adopted condition performs close to this relaxed model for thresholds from 10 to 40~dB, because the coverage is mainly limited by factors other than the tag-side SINR requirement, which depend on the tag direction. For tags in directions away from the user, the mutual interference can be effectively suppressed, so the coverage is limited by the incident power at long range. For tags in directions close to the user, the coverage is limited by the conflict between the user SINR and RFID power requirements. Covering tags at long range requires nearly the full power budget to meet the tag power sensitivity, whereas the user SINR requirement can be met with a small fraction of it. Hence, the sensing beam carries most of the transmit power and requires a large gain difference between the tag and user directions, which an array with a finite number of antennas cannot provide between close directions.

\begin{figure}[!t]
	\centering

	\includegraphics[width=1.0\columnwidth]{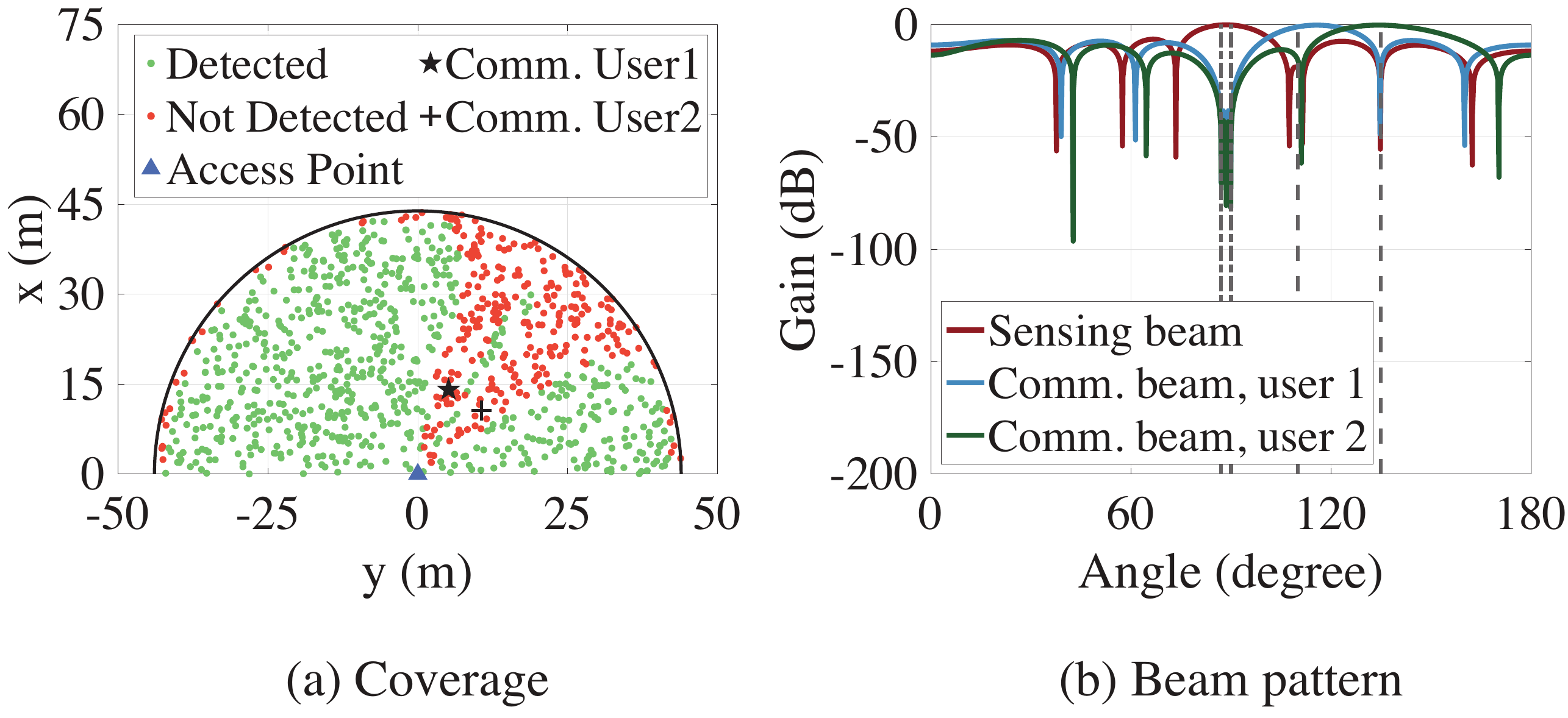}

	\caption{Performance of the proposed codebook with two communication users located at $110^\circ$ and $135^\circ$, each 15 meters from the access point. Subplot (a) shows the interrogation coverage for a $3^\circ$ sector interval, while subplot (b) shows the beamforming patterns for the $87^\circ$ -- $90^\circ$ sector. The beams suppress mutual interference while serving both users.}
	\label{fig:multi_tag_two_users}
\end{figure}

\begin{figure}[!t]
	\centering

	\includegraphics[width=1.0\columnwidth]{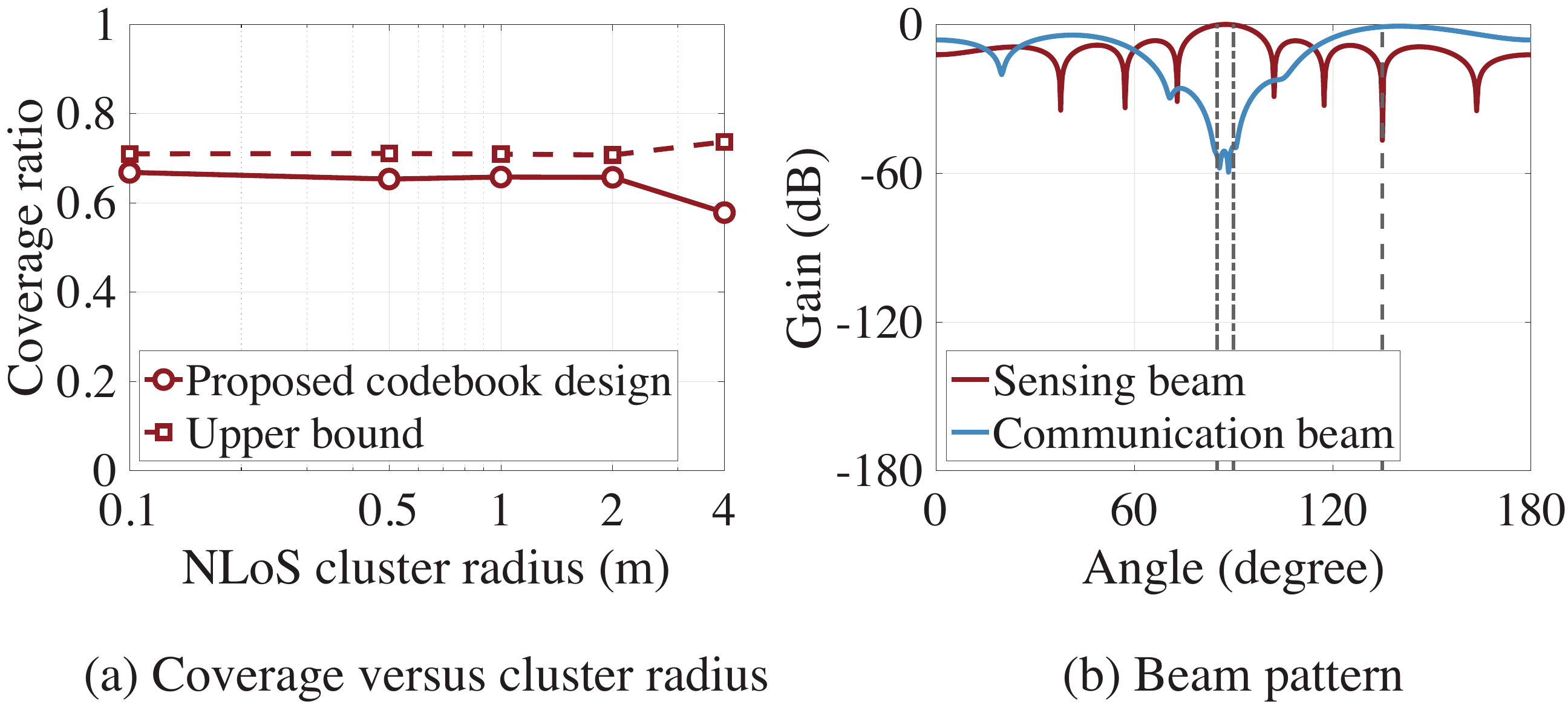}

	\caption{Performance under Rician multipath with a Rician factor of 7 dB and a $5^\circ$ sector interval. Subplot (a) shows average coverage versus NLoS cluster radius, while subplot (b) shows the beamforming patterns for a 2-meter cluster and the $85^\circ$ -- $90^\circ$ sector.}
	\label{fig:multi_tag_multipath}
\end{figure}

\textbf{How does the proposed codebook perform with multiple users and multipath channels?} We consider a two-user scenario in \figref{fig:multi_tag_two_users}, with users at $110^\circ$ and $135^\circ$, both 15 meters away, and a sector interval of $3^\circ$. \figref{fig:multi_tag_two_users}(a) shows that the proposed design maintains substantial interrogation coverage while satisfying both users’ SINR requirements, although stronger interference and spatial constraints produce more uninterrogated tags around and between their directions. The beamforming patterns in \figref{fig:multi_tag_two_users}(b) show that the sensing beam forms nulls toward both users, while each communication beam targets its intended user and suppresses interference over the sensing sector. We next evaluate the design under multipath propagation in \figref{fig:multi_tag_multipath}. The access-point-to-tag channel is modeled as a spatially correlated Rician channel,
\begin{equation}
	\bg_t =
	\rho_t
	\left(
		\sqrt{\frac{K_{\rm{R}}}{K_{\rm{R}}+1}}\,
		\ba(\theta_t)
		+
		\sqrt{\frac{1}{K_{\rm{R}}+1}}\,
		\bR_t^{1/2}\bz_t
	\right),
	\label{eq:rician_tag_channel}
\end{equation}
where $\rho_t$ is the complex large-scale channel coefficient, $K_{\rm{R}}$ is the Rician factor, $\ba(\theta_t)$ is the LoS array response, $\bz_t \sim \cC\cN(\bm{0}, \bI)$ is a complex Gaussian vector, and $\bR_t$ is the spatial correlation matrix of the non-line-of-sight (NLoS) component. We set $K_{\rm{R}}$ to 7 dB~\cite{3gpp_tr38901}. The NLoS component is generated by a local-scattering cluster centered at the tag, whose radius controls the angular spread. The access-point-to-user channel is LoS, whereas the tag-to-user channel follows an independent scalar Rician fading model with the same factor. 
Using a sector interval of $5^\circ$, \figref{fig:multi_tag_multipath}(a) shows that coverage remains nearly unchanged for cluster radii up to 2 meters and decreases at 4 meters, where the larger angular spread weakens the correspondence between the reference-tag and actual-tag channels. In contrast, the upper bound slightly increases, since a per-tag optimized beam can exploit the angularly spread NLoS paths, which reduce deep fades. The representative patterns in \figref{fig:multi_tag_multipath}(b), obtained with a cluster radius of 2 meters, show that the codebook continues to suppress mutual interference, demonstrating its applicability beyond the LoS setting.


\section{Conclusion} \label{sec:conclusion}
We investigated beamforming and codebook design for an ISAC system with backscattering RFID tags. For the single-tag scenario, we developed a zero-forcing method with optimized power allocation and a joint beamforming design based on transmit power minimization. The joint design achieved a larger interrogation distance and lower power consumption, especially under low communication user SINR requirements, whereas zero-forcing provided comparable performance under stricter requirements. For the multi-tag scenario, we developed a site-specific, sector-based codebook from an offline channel database at reference locations and solved the coverage problem using semidefinite relaxation and generalized Benders decomposition. Although designed on a finite reference grid, the codebook outperformed the benchmarks for randomly located off-grid tags, demonstrating its applicability throughout the region of interest without prior knowledge of tag locations or channels.
The proposed codebook framework can be extended in several directions. First, NLoS tags with blockage and richer multipath propagation may require robust designs that accommodate greater channel variation. Second, reference locations could be grouped by tag distribution and spatial channel characteristics rather than solely by angular sector \cite{Zhang2022}, potentially reducing codebook redundancy and beam-sweeping overhead. Finally, a real-world testbed \cite{Chen2016,McCormick2017} could evaluate the framework under practical channel, hardware, and protocol conditions.

\appendix

\subsection{Proof of Proposition 1} \label{appendix:proof_proposition_1}
This proof shows that the tag-side and reader-side SINR constraints in \eqref{eq:opt_PA_constraint_1} and \eqref{eq:opt_PA_constraint_2}, respectively, impose two lower bounds on the required sensing power. First, substituting the tag-side SINR expression in \eqref{eq:SINR_tag} into \eqref{eq:opt_PA_constraint_1} yields
\begin{subequations} \label{eq:opt_PA_constraint_1_rewrite}
	\begin{align}
		P_t^{(\rm{s})} & \geq \frac{\gamma_\rm{t}}{|\bg_t^H \bar{\bff}_t^{(\rm{s})}|^2} (\sum_{u=1}^{U}P_u^{(\rm{c})}|\bg_t^H \bar{\bff}_u^{(\rm{c})}|^2 + \sigma_\rm{t}^2) \\
		& = \frac{\gamma_\rm{t} \sigma_\rm{t}^2}{|\bg_t^H \bar{\bff}_t^{(\rm{s})}|^2}, \label{eq:opt_PA_constraint_1_rewrite_B}
	\end{align}
\end{subequations}
where the equality follows from the zero-forcing design introduced in \sref{subsec:ZF_beamforming}, under which every communication beam is nulled at the tag, i.e., $|\bg_t^H \bar{\bff}_u^{(\rm{c})}|=0, \, \forall u \in \mathcal{U}$. Similarly, substituting the reader SINR expression \eqref{eq:SINR_reader} into \eqref{eq:opt_PA_constraint_2} yields
\begin{align} \label{eq:opt_PA_constraint_2_rewrite}
	P_t^{(\rm{s})} \geq \frac{\gamma_\rm{r}}{\eta |\bw_t^H \bg_t|^2 |\bg_t^H \bar{\bff}_t^{(\rm{s})}|^2} (\eta \, \sigma_\rm{t}^2 |\bw_t^H \bg_t|^2 + \sigma_\rm{r}^2).
\end{align}
Observing \eqref{eq:opt_PA_constraint_1_rewrite} and \eqref{eq:opt_PA_constraint_2_rewrite}, we can see that both constraints impose a lower bound on the sensing power $P_t^{(\rm{s})}$. Consequently, the minimum sensing power required for successful interrogation is determined by the larger of the two lower bounds, which identifies whether the tag-side or reader-side SINR constraint is dominant.

\subsection{Proof of Proposition 2} \label{appendix:proof_proposition_2}
The proof relies on two key properties. First, to minimize the total transmit power, the optimal solution must satisfy the dominant RFID SINR constraint and user SINR constraints with equality. 
Second, when we use zero-forcing beamforming, the direct zero-forced terms are exactly zero.
We demonstrate this for the case where the tag-side constraint \eqref{eq:opt_PA_constraint_1} is dominant, and the same reasoning applies when \eqref{eq:opt_PA_constraint_2} is dominant. Using the first property, we rewrite the constraints \eqref{eq:opt_PA_constraint_1} and \eqref{eq:opt_PA_constraint_3} as shown in \eqref{eq:opt_PA_linear_sys_case1}. 
\begin{figure*} [t!]
	\centering
	\begin{equation} \label{eq:opt_PA_linear_sys_case1} 
		\begin{cases}
			P_t^{(\rm{s})} = \frac{\gamma_\rm{t} \left[ \sum_{u=1}^{U}P_u^{(\rm{c})}|\bg_t^H \bar{\bff}_u^{(\rm{c})}|^2 + \sigma_\rm{t}^2 \right]}{|\bg_t^H \bar{\bff}_t^{(\rm{s})}|^2}, \\
			P_u^{(\rm{c})} = \frac{\gamma_\rm{u} \left[ \sum_{l \neq u}^{U} P_l^{(\rm{c})} |\bh_u^H \bar{\bff}_l^{(\rm{c})}|^2 + P_t^{(\rm{s})} |\bh_u^H \bar{\bff}_t^{(\rm{s})}|^2 + \eta |h_{t,u}|^2 (P_t^{(\rm{s})} |\bg_t^H \bar{\bff}_t^{(\rm{s})}|^2 + \sum_{v=1}^{U} P_v^{(\rm{c})} |\bg_t^H \bar{\bff}_v^{(\rm{c})}|^2 + \sigma_\rm{t}^2) + \sigma_\rm{u}^2 \right]}{|\bh_u^H \bar{\bff}_u^{(\rm{c})}|^2}, \ \forall u \in \mathcal{U}.
		\end{cases}
	\end{equation} 
	\hrule
\end{figure*}
Then, the second property leads to the simplified set of equations in \eqref{eq:opt_PA_linear_sys_case1_simplified}.
\begin{equation} \label{eq:opt_PA_linear_sys_case1_simplified}
	\begin{cases}
		P_t^{(\rm{s})} = \frac{\gamma_\rm{t} \sigma_\rm{t}^2}{|\bg_t^H \bar{\bff}_t^{(\rm{s})}|^2}, \\
		P_u^{(\rm{c})} = \frac{\gamma_\rm{u} \left[ \eta |h_{t,u}|^2 (P_t^{(\rm{s})} |\bg_t^H \bar{\bff}_t^{(\rm{s})}|^2 + \sigma_\rm{t}^2) + \sigma_\rm{u}^2 \right]}{|\bh_u^H \bar{\bff}_u^{(\rm{c})}|^2}, \,\forall u \in \mathcal{U}.
	\end{cases}
\end{equation}
The first equation directly yields the optimal sensing power. Substituting it into the second equation gives the optimal communication powers. Summing these powers gives the minimum total transmit power required by the fixed zero-forcing directions. Comparing this value with the power budget verifies the feasibility condition stated in Proposition 2. If the required total power does not exceed the budget, the closed-form allocation is feasible and optimal.

\subsection{Connection Between the Sector-Wise Codebook Design and the Codebook Design in \eqref{eq:opt}}
\label{appendix:connection_opt_sector}

Let $\Omega$ denote the complete region of interest, which is partitioned
into $N$ nonoverlapping sectors such that
\begin{equation} 
    \Omega = \bigcup_{n=1}^{N}\mathcal{S}_n,
    \quad
    \mathcal{S}_n\cap\mathcal{S}_k=\emptyset,
    \quad n\neq k.
\end{equation}
The number of sectors equals the codebook cardinality $|\cF|=N$ defined in Section~\ref{sec:problem_formulation}, and the $n^{\rm{th}}$ codeword is assigned to sector $\mathcal{S}_n$. For a potential tag location $(r,\theta)\in\Omega$, define the interrogation-success indicator obtained when the SINRs are evaluated with the $n^{\rm{th}}$ codeword as
\begin{equation} 
    q_n(r,\theta)
    =
    \mathds{1}\left(
        \rm{SINR}_{r,\theta,n}^{(\rm{t})}
        \geq\gamma_\rm{t}
    \right)
    \mathds{1}\left(
        \rm{SINR}_{r,\theta,n}^{(\rm{r})}
        \geq\gamma_\rm{r}
    \right).
\end{equation}
Thus, the interrogation indicator of a tag located at $(r_t,\theta_t)$
in the original problem \eqref{eq:opt} can be written as $I_t=\max_{n\in\mathcal{N}}q_n(r_t,\theta_t)$.
Since the tag locations are unknown at the codebook design stage, we evaluate the design objective in expectation under a uniform spatial distribution. Accordingly, we model each tag location as uniformly distributed over the physical area of $\Omega$. Let $A_{\Omega}=\int_{\Omega} r\,dr\,d\theta$
denote the physical area of the region. Since the tags are uniformly
distributed with respect to physical area, their spatial probability
density is $1/A_{\Omega}$ with respect to the area element
$dA=r\,dr\,d\theta$. Using the linearity of expectation and the
definition
$I_t=\max_{n\in\mathcal{N}}q_n(r_t,\theta_t)$, we obtain
\begin{align} 
    \mathbb{E}\left[
        \sum_{t=1}^{T} I_t
    \right]
    &=
    \sum_{t=1}^{T}\mathbb{E}[I_t]
    =
    \sum_{t=1}^{T}
    \int_{\Omega}
        \max_{n\in\mathcal{N}}
        q_n(r,\theta)
        \frac{1}{A_{\Omega}}
        \,dA
    \nonumber\\
    &=
    \frac{T}{A_{\Omega}}
    \int_{\Omega}
        \max_{n\in\mathcal{N}}
        q_n(r,\theta)
        \,r\,dr\,d\theta.
    \label{eq:expected_original_coverage}
\end{align}
The integral in \eqref{eq:expected_original_coverage} represents the
physical area in which a tag can be successfully interrogated by at
least one codeword. Because $T/A_{\Omega}$ is independent of the
codebook design, maximizing the expected number of successfully
interrogated tags in \eqref{eq:opt} is equivalent to maximizing this
coverage area.
The proposed sector-wise design assigns the $n^{\rm{th}}$ codeword to
sector $\mathcal{S}_n$ and considers the following coverage:
\begin{equation} 
    C_{\rm{sector}}(\mathcal{F})
    =
    \sum_{n=1}^{N}
    \int_{\mathcal{S}_n}
        q_n(r,\theta)
        \,r\,dr\,d\theta.
    \label{eq:sector_coverage}
\end{equation}
For every $(r,\theta)\in\mathcal{S}_n$, we have 
\begin{equation} \label{eq:sector_coverage_lower_bound}
	\max_{k\in\mathcal{N}}q_k(r,\theta)\geq q_n(r,\theta).
\end{equation}
Integrating \eqref{eq:sector_coverage_lower_bound} over the sector partition gives 
\begin{align} 
    \int_{\Omega}
        \max_{k\in\mathcal{N}}
        q_k(r,\theta)
        \,r\,dr\,d\theta
	&=
    \sum_{n=1}^{N}
    \int_{\mathcal{S}_n}
        \max_{k\in\mathcal{N}}q_k(r,\theta)
        \,r\,dr\,d\theta
    \nonumber\\
    &\geq
    \sum_{n=1}^{N}
    \int_{\mathcal{S}_n}
        q_n(r,\theta)
        \,r\,dr\,d\theta.
    \label{eq:sector_lower_bound}
\end{align}
Hence, for any codebook $\cF$ that satisfies the communication user SINR and transmit power constraints, the sum of the sector-wise objectives
$C_{\rm{sector}}(\cF)$ is a lower bound on the true spatial coverage
objective of the original problem \eqref{eq:opt}. Problem
\eqref{eq:opt_BF} maximizes one term of \eqref{eq:sector_coverage}, and
solving it for all sectors constructs the complete codebook. The bound
becomes tight when every location covered by at least one codeword is
also covered by the codeword assigned to its sector.


\end{document}